\documentclass[12pt]{article}
\usepackage{epsfig}
\usepackage{amssymb}
\usepackage{amsmath}
\usepackage{amsfonts}
\usepackage{amsthm}
\usepackage{graphicx}
\usepackage{mathrsfs}
\usepackage[dvips]{color}
\usepackage{multirow}

\newcommand{\bsigma}{\boldsymbol{\sigma}}

\newcommand{\R}{\mathbb{R}}
\newcommand{\C}{\mathbb{C}}
\newcommand{\Z}{\mathbb{Z}}

\newcommand{\fa}{\mathfrak{a}}
\newcommand{\fb}{\mathfrak{b}}
\newcommand{\fc}{\mathfrak{c}}

\newcommand{\fh}{\mathfrak{h}}

\newcommand{\fn}{{\mathfrak{n}}}

\newcommand{\fs}{\mathfrak{s}}

\newcommand{\fz}{\mathfrak{z}}

\newcommand{\fK}{\mathfrak{K}}

\newcommand{\bk}{\mathbf{k}}

\newcommand{\bx}{\mathbf{x}}

\newcommand{\bE}{\mathbf{E}}

\newcommand{\bH}{\mathbf{H}}
\newcommand{\bI}{\mathbf{I}}

\newcommand{\bM}{\mathbf{M}}

\newcommand{\cA}{{\mathcal{A}}}

\newcommand{\cH}{\mathcal{H}}

\newcommand{\cK}{\mathcal{K}}

\newcommand{\cO}{\mathcal{O}}
\newcommand{\cP}{\mathcal{P}}

\newcommand{\cR}{\mathcal{R}}

\newcommand{\cT}{\mathcal{T}}
\newcommand{\cU}{\mathcal{U}}

\newcommand{\be}{\begin{equation}}
\newcommand{\ee}{\end{equation}}
\newcommand{\bea}{\begin{eqnarray}}
\newcommand{\eea}{\end{eqnarray}}
\newcommand{\nn}{\nonumber}

\newcommand{\ed}{\end{document}}

\newcommand{\bi}{\begin{itemize}}
\newcommand{\ei}{\end{itemize}}

\newcommand{\bce}{\begin{center}}
\newcommand{\ece}{\end{center}}

\newcommand{\sS}{\mathscr{S}}
\newcommand{\sT}{\mathscr{T}}

\newcommand{\RE}{{\rm Re}}

\newcommand{\bPhi}{{\boldsymbol{\Phi}}}

\newcommand{\bcK}{{\boldsymbol{\cK}}}

\newcommand{\bcH}{{\boldsymbol{\cH}}}
\newcommand{\bcU}{{\boldsymbol{\cU}}}

\newcommand{\for}{{\mbox{\rm for}}}

\newcommand{\cx}{{\check{x}}}

\begin{document}
 
\title{Low-Frequency Scattering of TE and TM Waves by an Inhomogeneous Medium with Planar Symmetry}

\author{Farhang Loran\thanks{E-mail address: loran@iut.ac.ir},
~ Ali~Mostafazadeh\thanks{Corresponding author, E-mail address:
amostafazadeh@ku.edu.tr},~\ and Cem Yeti\c{s}mi\c{s}o\u{g}lu\thanks{E-mail address: cem.yetismisoglu@medipol.edu.tr}\\[6pt]
$^*$Department of Physics, Isfahan University of Technology, \\ Isfahan 84156-83111, Iran\\[6pt]
$^\dagger$Departments of Mathematics and Physics, Ko\c{c} University,\\  34450 Sar{\i}yer, Istanbul, T\"urkiye \\[6pt]
$^\ddagger$School of Engineering and Natural Sciences, Istanbul Medipol University,\\ 34810 Beykoz, Istanbul, T\"urkiye}

\date{ }
\maketitle

\begin{abstract}
Stationary scattering of TE and TM waves propagating in an isotropic medium with planar symmetry is described by Bergmann's equation in one dimension. This is a generalization of Helmholtz equation which allows for developing transfer matrix methods to deal with the corresponding scattering problems. We use a dynamical formulation of stationary scattering to study the low-frequency scattering of these waves when the inhomogeneities of the medium causing the scattering are confined to a planar slab. This formulation relies on the construction of an effective two-level non-Hermitian quantum system whose time-evolution operator determines the transfer matrix. We use it to construct the low-frequency expansions of the transfer matrix and the reflection and transmission coefficients of the medium, introduce a generalization of Brewster's angle for inhomogeneous slabs at low frequencies, and derive analytic conditions for transparency and reflectionlessness of $\cP\cT$-symmetric and non-$\cP\cT$-symmetric slabs at these frequencies. We also discuss the application of this method to deal with the low-frequency scattering of TE and TM waves when the carrier medium occupies a half-space and the waves satisfy boundary conditions with planar symmetry at the boundary of the half-space. Because acoustic waves propagating in a compressible fluid with planar symmetry are also described by Bergmann's equation, our results apply to the low-frequency scattering of these waves.

\end{abstract}

\section{Introduction}
\label{s1}

The study of scattering of waves involves understanding how they interact with objects or irregularities in the carrier medium and provides valuable information into the material properties and structural characteristics of the medium. The principal examples are the scattering of electromagnetic and acoustic waves which have important applications across physics, engineering, and other areas of applied science. Their utility in probing matter non-invasively makes them into a powerful analysis tool with a wide range of applications including radars, sonars, and tomography. There are many instances where the attenuation of the waves limit the effectiveness of these applications unless their frequency is sufficiently small. {Principle examples are long-range and under-water communications \cite{Bernstein-1974,Bannister-1984,Kaushal-2016,Smith-2019}, remote sensing  \cite{deBadereau-2003,Omer-2020}, and atmospheric physics  \cite{Cummer-2020,Hayakawa-2020,Hartinger-2022}.}  {The broad range of its applications has motivated the study of low-frequency scattering of electromagnetic \cite{Stevenson-1953} and acoustic waves \cite{Kriegsmann-1983,Sjoberg-2009}. For a general review and textbook treatments of the subject, see \cite{Kleinman-1986} and \cite{newton-book,dassios-book}, respectively. A mathematically rigorous discussion providing  references to earlier mathematical literature is provided in \cite{Ammari-2000}.}

If the carrier medium has planar symmetry and the scattering arises due to its inhomogeneities, the scattering problem is effectively one-dimensional. For example, consider the scattering of time-harmonic electromagnetic waves {by the inhomogeneities of a charge-free, linear, and isotropic carrier medium $\sS$ with planar symmetry.} To study this phenomenon, we choose a Cartesian coordinate system in which $\sS$ exhibits translational symmetry along the $y$ and $z$ directions. Then its permittivity and permeability depend only on the $x$ coordinate, and we can respectively denote them by $\varepsilon(x)$ and $\mu(x)$.\footnote{In general, the permittivity and permeability of $\sS$ also depend on the wavenumber of the incident wave. Since this dependence does not affect our analysis, we do not make it explicit.} Let $\hat{\varepsilon}(x) = \varepsilon(x)/\varepsilon_0$ and $\hat{\mu}(x) = \mu(x)/\mu_0$ label their respective relative values with respect to the vacuum or a homogeneous background,  i.e., 
	\begin{align}
	&\varepsilon_0:=\lim_{x\to\pm\infty}\varepsilon(x),
	&&\mu_0:=\lim_{x\to\pm\infty}\mu(x). 
	\nn
	\end{align}
	
The purpose of the present article is to provide a systematic method for the study of the low-frequency scattering of time-harmonic electromagnetic waves due to the inhomogeneities of $\sS$ under the following conditions.
	\begin{enumerate}
	\item  The inhomogeneities of $\sS$ are confined to a planar slab of thickness $\ell$ positioned between the planes given by $x=a$ and $x=a+\ell$, as shown in Fig.~\ref{fig1}. This means that 
	\be
	\hat{\varepsilon}(x)=\hat{\mu}(x)=1\quad\for\quad x\notin[a,a+\ell].
	\label{condi1}
	\ee%
	\begin{figure}
	\begin{center}
        \includegraphics[scale=.27]{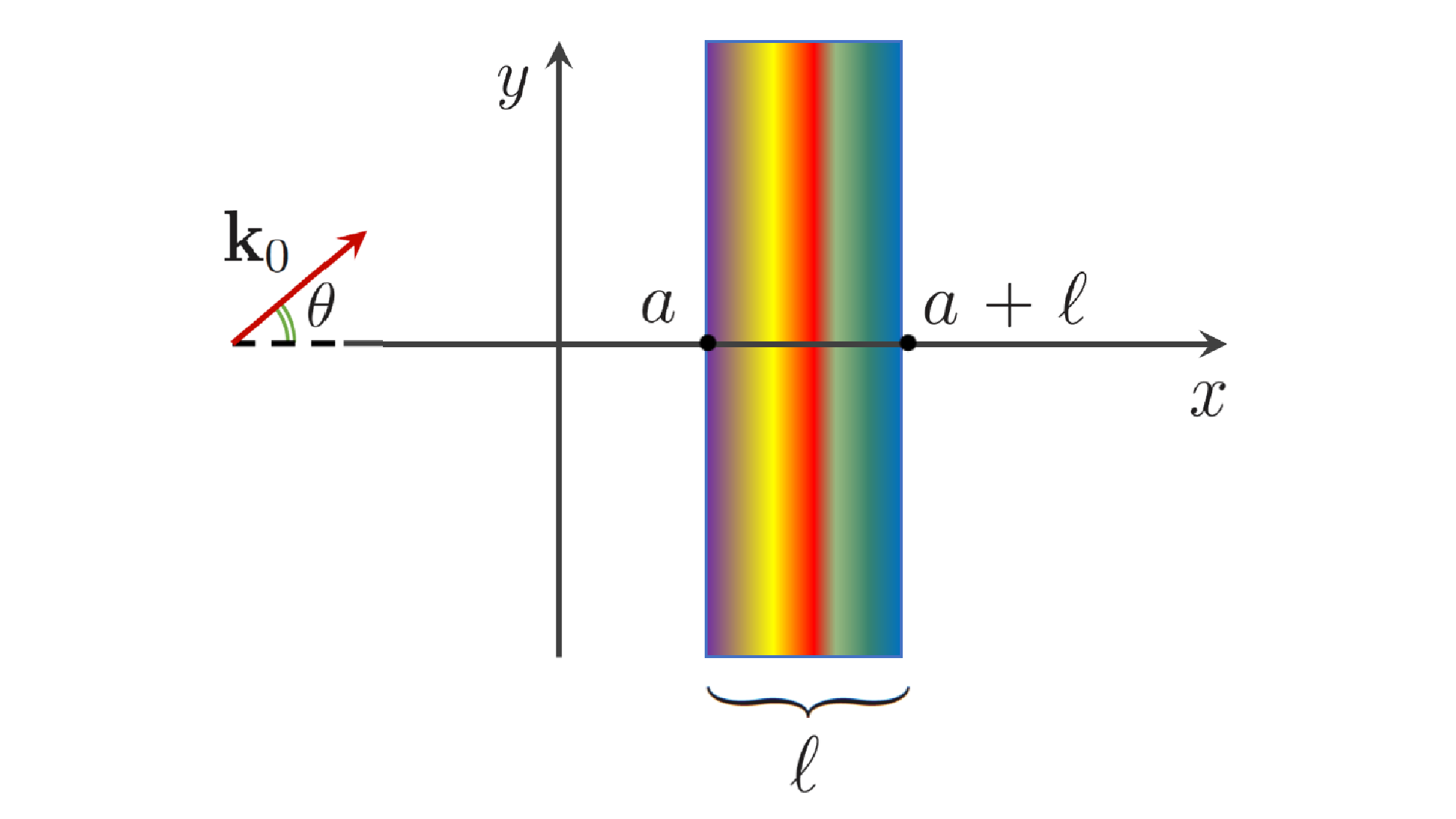}\hspace{-1.5cm}
        \includegraphics[scale=.27]{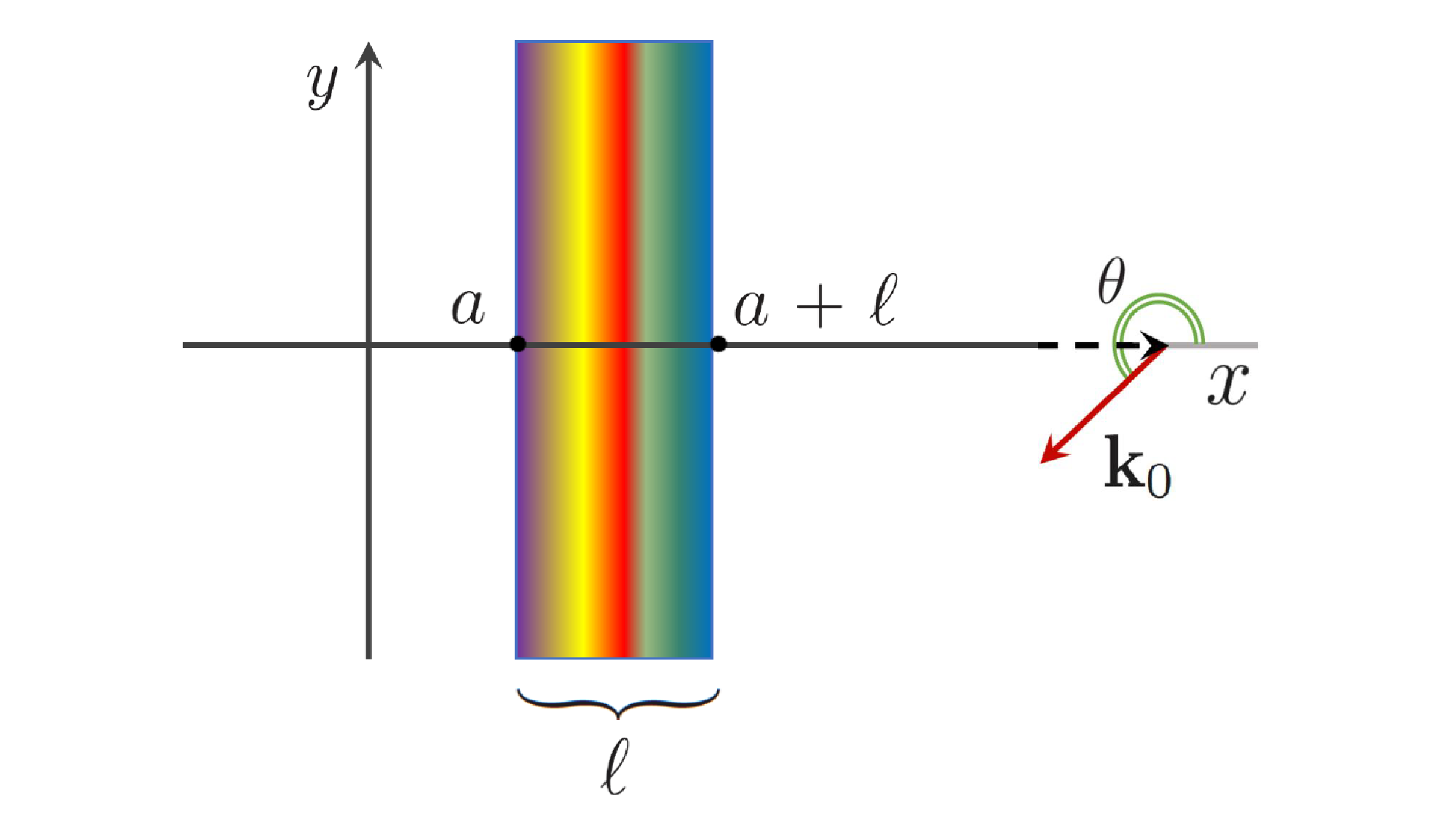} 
        \caption{Schematic views of the slab containing the inhomogeneities of the medium that cause the scattering. The incident wave vector $\bk_0$ is shown as a red arrow for a left-incident wave (on the left) and a right-incident wave (on the right). The incidence angles $\theta$ for left-incident and right-incident waves respectively satisfy $-90^\circ<\theta<90^\circ$ and $90^\circ<\theta<270^\circ$.}
        \label{fig1}
        \end{center}
        \end{figure}%
	\item The relative permittivity and permeability of the medium are bounded away from zero, i.e., there are positive real numbers $b_\pm$ such that for all $x\in[a,a+\ell]$,
	\begin{equation}
	0<{b_-}\leq|\hat{\varepsilon}(x)|\leq{b_+}, \qquad \qquad 0<{b_-}\leq|\hat{\mu}(x)|\leq{b_+}. 
	\label{bounds}
	\end{equation}
	\end{enumerate}
Here by ``low-frequency scattering,'' we mean the scattering of planewaves whose wavenumber $k$ is much smaller than $\ell^{-1}$, i.e., $k\ell\ll 1$.

It is well-known that the scattering problem for electromagnetic waves by the inhomogeneities of the medium $\sS$ reduces to that of its transverse electric (TE) and transverse magnetic (TM) modes \cite{griffiths-book}. In dealing with time-harmonic TE (respectively TM) waves, we orient the $z$-axis of our coordinate system along the electric field $\bE$ (respectively magnetic field $\bH$). Then the incident wavevector $\bk_0$ lies in the $x$-$y$ plane, and we have
	\begin{align}
	&\bE =E_0\, e^{i(k_y y-\omega t)}\phi(x)\,\hat{\textbf{e}}_z \quad \:\mbox{for TE waves}, \label{Ez=}\\
	&\bH =H_0\, e^{i(k_y y-\omega t)}\phi(x)\,\hat{\textbf{e}}_z \quad \mbox{for TM waves},\label{Hz=}\\
    	&\text{\textbf{k}}_0 = k_x \hat{\text{\textbf{e}}}_x + k_y \hat{\text{\textbf{e}}}_y, 
	\quad \quad k_x := k \cos \theta, 
	\quad \quad  k_y := k \sin \theta,
	\end{align}
where $E_0$ and $H_0$ are constant amplitudes, $\omega$ and $\theta$ are respectively  the angular frequency and incidence angle of the incident wave, $k$ is its wavenumber, $\hat{\text{\textbf{e}}}_u$ denotes the unit vector along the positive $u$ axis, $u\in\{x,y,z\}$, and $\phi$ is a possibly complex-valued function that describes the spatial profile of the wave. 

{For $k_x>0$ (respectively $k_x<0$)}, $-90^\circ<\theta<90^\circ$ (respectively $90^\circ<\theta<270^\circ$), and we refer to the incident wave as a ``left-incident wave'' (respectively ``right-incident wave"). See Fig.~\ref{fig1}.

Imposing Maxwell's equations on the ansatzes \eqref{Ez=} and \eqref{Hz=}, we find \cite{ptep-2024b}
	\begin{align}
	&\mathbf{H}= \frac{E_0\,e^{i(k_y y-\omega t)}}{c\, \mu(x)}
	\left[\sin \theta\, \phi(x) \hat{\mathbf{e}}_x+ 
	\mbox{\large$\frac{i}{k}$}\, \partial_x \phi(x) \hat{\mathbf{e}}_y\right]
	\quad \mbox{~~for TE waves}, 
	\label{TE-H=}\\
	&\mathbf{E}= -\frac{H_0\,e^{i(k_y y-\omega t)}}{c\, \varepsilon(x)}
	\left[\sin \theta\, \phi(x) \hat{\mathbf{e}}_x+ 
	\mbox{\large$\frac{i}{k}$}\, \partial_x \phi(x) \hat{\mathbf{e}}_y\right] \quad 
	\mbox{for TM waves},
	\label{TM-E=}\\
	&\alpha(x)\, \partial_x\left[\alpha(x)^{-1}\partial_x\phi(x)\right] + \mathfrak{K}^2\,
	\tilde{\mathfrak{n}}(x)^2 \phi(x) = 0, 
	\label{Bergmann}
	\end{align}
where 
	\begin{align}
	&c:=\frac{\omega}{k},
	&&\alpha :=\left\{\begin{array}{ccc}
	\hat{\mu} & \text{for} & \text{TE waves, } \\
	\hat{\varepsilon} & \text{for} & \text{TM waves, }
	\end{array}\right. 
	\label{alpha=}\\
	&\mathfrak{K} := \vert k_x \vert = k \vert \cos \theta \vert, 
	&&\tilde{\mathfrak{n}}(x) := \pm 	\vert \sec \theta 
	\vert \sqrt{\mathfrak{n}(x)^2 - \sin^2 \theta}, 
	\label{kneqn}
	\end{align}
$\fn(x)$ is the complex refractive index of the medium, which satisfies $\fn(x)^2:=\hat{\varepsilon}(x) \hat{\mu}(x)$, and the sign in \eqref{kneqn} stands for the sign of the real part of $\fn(x)$; it is positive for ordinary materials and negative for the negative-index metamaterials \cite{Veselago,Pendry-2008,ap-2016}. 

The above analysis reduces the stationary scattering of the TE and TM waves by the inhomogeneities of the medium $\sS$ to the one defined by the wave equation (\ref{Bergmann}). This is known as the Bergmann's equation in acoustics, where it is used to model the propagation of sound waves propagating in a compressible fluid with planar symmetry \cite{Bergmann,Martin}.\footnote{In acoustic applications, $\alpha(x)$ represents the density of the fluid, $\theta=0$, $\tilde\fn(x)=\fn(x):=\fc_0/\fc(x)$ where $\fc(x)$ is the speed of sound in the fluid, and $\fc_0$ is its value at spatial infinity \cite{Bergmann,Martin}. Some authors call $\fc_0/\fc(x)$ the ``refractive index'' of the fluid \cite{colton-kress}.}

{If $\sS$ is made of nonmagnetic material, $\alpha=\hat\mu=1$, and (\ref{Bergmann}) reduces to the Helmholtz equation, 
	\be
	\partial_x^2\phi(x)+ \mathfrak{K}^2\,\tilde{\mathfrak{n}}(x)^2 \phi(x) = 0.
	\label{HH-eq}
	\ee}%
Because this equation is identical to the time-independent Schr\"odinger equation, 
	\[-\partial_x^2\phi(x)+v(x)\phi(x)=\fK^2\phi(x),\] 
for the finite-range potential,
	\begin{equation}
	v(x)=\begin{cases} \fK^2(1-\tilde{\fn}^2) & \text { for } x \in [a,a+\ell], \\ 
	0 & \text { for } x \notin [a,a+\ell], \end{cases} 
	\label{potential}
	\end{equation}
one can use the tools of potential scattering to address this scattering problem. {This has led to the development of analytic and semi-analytic methods based on the semi-classical (WKB) approximation \cite{Bremmer-1951,Berk-1967}, perturbative and iterative solutions of certain variants of the Lippmann-Schwinger equation \cite{Hassab-1972,Chen-1978}, and high-order Born approximation \cite{Reboucas-2022}. Ref.~\cite{Bolle-1985} provides a mathematically rigorous but practically cumbersome treatment of low-frequency potential scattering in one dimension.}  Ref.~\cite{jpa-2021} employs the standard transfer matrix of the potential scattering \cite{sanchez,tjp-2020} and its Dyson series expansion \cite{ap-2014} to develop an effective scheme for constructing the low-frequency expansion of the scattering data {associated with the Helmholtz equation \eqref{HH-eq}.} 

At this stage, it is tempting to map Bergmann's equation to a Helmholtz equation by a change of the dependent variable and try to use the results of Ref.~\cite{jpa-2021} to obtain the low-frequency expansions of the scattering data for general TE and TM waves and magnetic material. Because the resulting Helmholtz equation involves $\partial_x\alpha$, as noted in Ref.~\cite{Martin}, this approach leads to serious difficulties in many practical situations where $\hat\varepsilon$ or $\hat\mu$ and consequently $\alpha$ are discontinuous functions. This problem does not arise for Bergmann's equation~(\ref{Bergmann}), because imposing Maxwell's boundary conditions at the discontinuities of $\alpha$, we find that $\phi$ and $\alpha^{-1}\partial_x\phi$ are continuous functions of $x$, \cite{ptep-2024b}. The standard transfer matrix methods used in potential scattering cannot be directly applied for the scattering problem defined by the Bergmann's equation, because in potential scattering it is $\partial_x\phi$ and not $\alpha^{-1}\partial_x\phi$ that is required to be continuous.

Recently, two of us have introduced an alternative transfer matrix method to deal with the stationary scattering defined by the Bergmann's equation (\ref{Bergmann}) and showed that, similarly to the transfer matrix of potential scattering, it admits a Dyson series expansion generated by a non-Hermitian two-level  Hamiltonian \cite{ptep-2024b}. In the present article, we explore the utility of this expansion in the study of low-frequency scattering of general TE and TM waves for cases where $\sS$ need not be made of a nonmagnetic material. {In contrast to the approaches pursued in the literature on the subject, this allows for the derivation of explicit analytic formulas for the leading- and next-to-leading-order terms in the low-frequency expansions of the scattering date.} Because of the application of Bergmann's equation in acoustic \cite{Martin}, our results directly apply to the scattering of sound waves in effectively one-dimensional compressible fluids with space-dependent density and/or compressibility (speed of sound).  	

The organization of this article is as follows. In Section \ref{s2}, we use a variation of the approach of Ref.~\cite{ptep-2024b} to obtain a Dyson series expansion for the transfer matrix of the scattering defined by the Bergmann's equation. In Section \ref{s3}, we use the latter to derive the low-frequency expansion of the transfer matrix. In Section \ref{s4}, we {construct the low-frequency expansions of the reflection and transmission amplitudes for this scattering problem and discuss some of their physical consequences.} In Section \ref{s5}, we extend our results to the scattering problems defined by the Bergmann's equation in a half-space. These correspond to TE, TM, or sound waves propagating in the half-space, $x>0$, and subject to certain translationally-invariant boundary conditions on the boundary plane $x=0$. In particular, we obtain the low-frequency expansion of the reflection coefficient of the medium for general boundary conditions of Robin type. In Section \ref{s6}, we summarize our findings and present our concluding remarks.

\section{Transfer Matrix and Its Dyson Series Expansion}
\label{s2}

According to (\ref{condi1}), for $x\notin [a,a+\ell]$, $\alpha(x)=\fn(x)^2=\tilde\fn(x)^2=1$, and the Bergmann's equation (\ref{Bergmann}) reduces to the Helmholtz equation $\partial_x^2\phi(x)+\fK^2\phi(x)=0$. This implies that every solution of (\ref{Bergmann}) fulfills
	\begin{equation}
	\phi(x)= \begin{cases}A_{-} e^{i\mathfrak{K}x}+
	B_{-} e^{-i\mathfrak{K}x} & \text { for } x \leqslant a, \\ 
	A_{+} e^{i\mathfrak{K} x}+
	B_{+} e^{-i\mathfrak{K}x} & \text { for } x \geqslant a+\ell,	
	\end{cases} 
	\label{asympsolns}
	\end{equation}
where $A_\pm$ and $B_\pm$ are possibly $k$-dependent complex parameters respectively representing the amplitudes of the right-going and left-going waves along the $x$ axis. The transfer matrix associated with the medium $\sS$ is the unique $k$-dependent $2\times 2$ matrix $\bM$ that satisfies
	\begin{equation}
	\left[\begin{array}{c} A_+ \\ B_+ \end{array}\right] = 
	\bM	\left[\begin{array}{c} A_- \\ B_- \end{array}  \right] , 			
	\label{M-def} 
	\end{equation}
and is independent of $A_\pm$ and $B_\pm$.

The scattering solutions, $\phi_l$ and $\phi_r$, of the Bergmann's equation that respectively describe the left-incident and right-incident waves satisfy 
	\begin{align}
	&\phi_l(x)=N_l\times\left\{\begin{array}{ccc}
	e^{i\mathfrak{K}x}+R^l\, e^{-i\mathfrak{K}x} & \for &
	x\leqslant a, \\ 
	T^l\, e^{i\mathfrak{K} x}&\for& x \geqslant a+\ell,	
	\end{array}\right.
	\label{left-inc}\\
	&\phi_r(x)=N_r\times\left\{\begin{array}{ccc}
	T^r\, e^{-i\mathfrak{K}x} & \for &
	x\leqslant a, \\ 
	e^{-i\fK x}+R^r\, e^{i\mathfrak{K} x}&\for& x \geqslant a+\ell,	
	\end{array}\right.
	\label{right-inc}
	\end{align}
where $N_{l/r}$ are the complex amplitudes of the left/right-incident waves, and $R^{l/r}$ and $T^{l/r}$ are respectively the corresponding complex reflection and transmission amplitudes. In view of (\ref{M-def}) -- (\ref{right-inc}), 
	\begin{align}
	&R^l = - \frac{M_{21}}{M_{22}}, 
	&& R^r= \frac{M_{12}}{M_{22}}, 
	\label{Rs=}\\
	& T^l=\frac{\det\text{\textbf{M}}}{M_{22}}, 
	&&T^r=\frac{1}{M_{22}},  
	\label{Ts=}
	\end{align}
where $M_{ab}$ denote the entries of $\bM$. 

The above definition of the transfer matrix, the reflection and transmission amplitudes, and their relationship given by (\ref{Rs=}) and \eqref{Ts=} are identical to the ones for potential scattering~\cite{sanchez,tjp-2020}. The only difference is that the wave equation we must use to compute the transfer matrix and the reflection and transmission amplitudes is the Bergmann's equation~\eqref{Bergmann}. Introducing the two-component wave function,
	\begin{equation}
	\mathbf{\Phi}(x):= \frac{1}{2}
	\left[\begin{array}{c}\phi(x)-i(\mathfrak{K}	
	\alpha)^{-1}\partial_{x}\phi(x)\\[3pt]
	\phi(x)+i(\mathfrak{K}\alpha)^{-1}\partial_{x}\phi(x)\end{array}\right], 	
	\label{statevector}
	\end{equation}
we can write this equation in the form of a time-dependent Schr\"odinger equation,
	\begin{equation}
	i\partial_x \mathbf{\Phi}(x) = \bcH (x) \mathbf{\Phi}(x),
	\label{2stateschrodinger}
	\end{equation}
for the non-Hermitian matrix Hamiltonian,
	\begin{equation}
	\bcH(x) :=\frac{\mathfrak{K}}{2}
	\begin{bmatrix} -\fh_+(x) & -\fh_-(x) \\ \fh_-(x) & \fh_+(x) \end{bmatrix}
	=-\frac{\mathfrak{K}}{2}
	\left[\frac{\tilde{\mathfrak{n}}(x)^2}{\alpha(x)}\,\bcK+\alpha(x)\,\bcK^T\right],
    	\label{Hamiltonian}
	\end{equation}
where $x$ plays the role of time, 
	\begin{align}
	&\fh_\pm:=\frac{\tilde{\mathfrak{n}}^2}{\alpha}\pm\alpha,
	&&\bcK:=\left[\begin{array}{cc}
	1&1\\
	-1&-1\end{array}\right]=i \bsigma_2+\bsigma_3,
	\label{bcK-def}
	\end{align}
$\bcK^T$ denotes the transpose of $\bcK$, and $\bsigma_j$'s are the Pauli matrices;
	\begin{align}
	&\bsigma_1 := \begin{bmatrix} 0 & 1 \\ 1 & 0 \end{bmatrix}, 
	&&\bsigma_2 := \begin{bmatrix} 0 & -i \\ i & 0 \end{bmatrix}, 
	&&\bsigma_3 := \begin{bmatrix} 1 & 0 \\ 0 & -1 \end{bmatrix}.
	\label{Pauli}
	\end{align}
	
An important property of $\bPhi(x)$, which follows from \eqref{asympsolns} and \eqref{statevector}, is that it satisfies
	\begin{align}
	&\mathbf{\Phi}(a) = e^{i\mathfrak{K}a \bsigma_3} 
	\left[\begin{array}{c} A_- \\ B_-    \end{array}   \right], 
	&& \mathbf{\Phi}(a+\ell) = e^{i\mathfrak{K}(a+\ell)\bsigma_3} \left[\begin{array}{c} A_+ \\ B_+\end{array}   \right].\nn
	\end{align}
By virtue of \eqref{M-def}, these equations imply 
	\be
	\mathbf{\Phi}(a+\ell)=e^{i\mathfrak{K}(a+\ell)\bsigma_3}\bM\,
	e^{-i\mathfrak{K}a \bsigma_3}\mathbf{\Phi}(a).
	\label{this-eq1}
	\ee
Because the evolution operator $\bcU(x,x_0)$ for the Hamiltonian (\ref{Hamiltonian}) evolves $\mathbf{\Phi}(x_0)$ into $\mathbf{\Phi}(x)$,  \eqref{this-eq1} suggests
	\be
	\bM=e^{-i\mathfrak{K}(a+\ell)\bsigma_3}
	\bcU(a+\ell,a)e^{i\mathfrak{K}a \bsigma_3}.
	\label{M=EUE}
	\ee
	
{Because $\hat\varepsilon$ and $\hat\mu$ are generally discontinuous functions at the boundaries of the slab, i.e., $x=a$ and $x=a+\ell$, the same applies to $\alpha$, $\tilde\fn$, and the entries of the effective Hamiltonian $\bcH$. Yet, in light of \eqref{condi1}, \eqref{bounds}, and \eqref{2stateschrodinger}, the two-component wave function $\bPhi$ must be continuous throughout $\R$, and in particular at the boundary points: $x=a$ and $x=a+\ell$. It is easy to see from \eqref{statevector} that this requirement is equivalent to the boundary conditions of our scattering problem, namely that $\phi(x)$ and $\alpha^{-1}\partial_x\phi$ are continuous at $x=a$ and $x=a+\ell$. For this reason, the use of the transfer matrix \eqref{M=EUE} for the purpose of solving the scattering problem of our interest is consistent with the boundary conditions of this problem.}

Next, we recall that for all $x_0\in\R$, the time-evolution operator $\bcU(x,x_0)$ is the unique solution of the initial-value problem,
	\begin{align}
	&i\partial_x\bcU(x,x_0) = \bcH(x)\,\bcU(x,x_0),
	&&\bcU(x_0,x_0)=\bI,\nn 
	\end{align}
where $\bI$ stands for the $2\times 2$ identity matrix. It admits the following Dyson series expansion \cite{dewitt2003global}.
	\begin{align}
	\bcU(x,x_0) &= \sT \exp \left[-i \int_{x_0}^x\!\!dx\, \bcH(x) \right] \nonumber\\
	&= \bI+ \sum_{n=1}^\infty (-i)^n \int_{x_0}^x \!\! dx_n
	\int_{x_0}^{x_n}\!\!dx_{n-1}\cdots \int_{x_0}^{x_2}\!\! dx_1\,
	\bcH(x_n)\bcH(x_{n-1})\cdots \bcH(x_1),
	\label{dyson}
	\end{align}
where $\sT$ denotes the ``time-ordering'' operation. Substituting this equation in (\ref{M=EUE}), we obtain a Dyson series expansion for the transfer matrix. 
	
Because the Hamiltonian (\ref{Hamiltonian}) is traceless, its time-evolution operator $\bcU(x,x_0)$ has unit determinant. In light of the fact that for all $\varphi\in\R$, $\det(e^{-i\varphi\bsigma_3})=1$, this observation together with (\ref{M=EUE}) imply $\det\bM=1$. Substituting this in (\ref{Ts=}) yields the transmission reciprocity, $T^l=T^r$. In the following we use $T$ to refer to both left and right transmission amplitudes. In particular, (\ref{Ts=}) becomes
	\be
	T=\frac{1}{M_{22}}.
	\label{T=}
	\ee
	
As a simple example of the application of (\ref{M=EUE}), consider the cases where our slab is made of a homogeneous material, i.e., inside the slab $\hat\varepsilon$ and $\hat\mu$ do not depend on $x$. Then the same applies to $\alpha$ and $\tilde\fn$. Consequently, for $a\leq x\leq a+\ell$, the Hamiltonian (\ref{Hamiltonian}) is $x$-independent, and \eqref{dyson} gives $\cU(a+\ell,a)=e^{-i\ell\bcH}$. Because $\bcH$ is a $2\times 2$ matrix, we can easily calculate $e^{-i\ell\bcH}$. Substituting the result for $\cU(a+\ell,a)$ in (\ref{M=EUE}), we find {the following expression for the transfer matrix of a homogeneous slab.}
	\begin{align}
	&\bM=\left[\begin{array}{cc}
	(\fc+i\tilde\fn_+\fs)e^{-i\fK\ell} & i\tilde\fn_-\fs\,e^{-i\fK(2a+\ell)}\\[6pt]
	 -i\tilde\fn_-\fs\,e^{i\fK(2a+\ell)} & (\fc-i\tilde\fn_+\fs)e^{i\fK\ell}\end{array}\right],
	 \label{M-homogen}
	 \end{align} 	 
where
	\begin{align}
	&\fc:=\cos(\fK\ell\tilde\fn),
	&&\fs=\sin(\fK\ell\tilde\fn),
	&&\tilde\fn_\pm:=
	\frac{1}{2}\left(\frac{\tilde\fn}{\alpha}\pm\frac{\alpha}{\tilde\fn}\right).
	\label{cs=}
	\end{align}
For TE waves with $a=0$, Eq.~\eqref{M-homogen} coincides with Eq.~(27) of Ref.~\cite{ap-2016} which uses a variant of the standard approach of imposing Maxwell's boundary conditions at the slab's boundaries to derive it.  
	
According to \eqref{Rs=} and  \eqref{T=} --  \eqref{cs=}, the reflection and transmission amplitudes of the homogeneous slab are given by
	\begin{align}
	&R^l=e^{2i(2a+\ell)\fK}R^r=
	\frac{i\tilde\fn_- e^{2ia\fK}}{\cot(\fK\ell\tilde\fn)+i\tilde\fn_+},
	\label{RR-homogen}\\
	&T=\frac{e^{-i\fK\ell}}{\cos(\fK\ell\tilde\fn)-i\tilde\fn_+\sin(\fK\ell\tilde\fn)}.
	\label{T-homogen}
	\end{align}

\section{Low Frequency Expansion of the Transfer Matrix}
\label{s3}

Consider cases where $a=0$. Then in view of \eqref{M=EUE} and \eqref{dyson}, the Dyson series expansion of the transfer matrix takes the form
	\begin{align}
	\bM &= e^{-i\mathfrak{K}\ell\bsigma_3}\Big[\bI+ 
	\sum_{n=1}^\infty (-i)^n \int_{0}^\ell \!\! dx_n
	\int_{0}^{x_n}\!\!dx_{n-1}\cdots \int_{0}^{x_2}\!\! dx_1\,
	\bcH(x_n)\bcH(x_{n-1})\cdots \bcH(x_1)\Big].
	\label{dyson-M}
	\end{align}
Changing the variables of integrations $x_j$ to $\check x_j:=x_j/\ell$ and noting that, according to (\ref{Hamiltonian}), $\bcH(x_j)$ is proportional to $\fK$, which equals $k|\!\cos\theta|$, we observe that the right-hand side of (\ref{dyson-M}) is a power series in $k\ell$. This shows that we can determine the low-frequency expansion of $\bM$ by finding a systematic method of computing the coefficients of this power series. 

First, we note that in view of (\ref{Hamiltonian}), $\bcH(x):=\frac{1}{2}[\fa(x)\bcK+\fb(x)\bcK^T]$, where 
	\begin{align}
	&\fa(x):=-\frac{\fK\,\tilde\fn(x)^2}{\alpha(x)}=
	-\frac{k\,\nu(x)}{|\!\cos\theta|}, \\
	&\fb(x):=-\fK\,\alpha(x)=-k|\!\cos\theta|\alpha(x), \\
	&\nu(x):=\frac{\fn(x)^2-\sin^2\theta}{\alpha(x)}=
	\frac{\varepsilon(x)\mu(x)-\sin^2\theta}{\alpha(x)}.
	\label{nu-def}
	\end{align}
This allows us to use the lemma given in Sec.~3 of Ref.~\cite{jpa-2021} to calculate the products of $\bcH(x_j)$'s appearing on the right-hand side of \eqref{dyson-M}. The result is 
	\begin{align}
	&\bcH(x_{2n})\bcH(x_{2n-1})\cdots \bcH(x_1)=
	\frac{k^{2n}}{2}
	\Big[c_{2n}(\bx_{2n})\bsigma_-+d_{2n}(\bx_{2n})\bsigma_+\Big],
	\label{H-even}
	\\
	&\bcH(x_{2n+1})\bcH(x_{2n})\cdots \bcH(x_1)=
	-\frac{k^{2n+1}}{2|\!\cos\theta|}
	\Big[d_{2n+1}(\bx_{2n+1})\bcK+\cos^2\theta\,
	c_{2n+1}(\bx_{2n+1})\bcK^T\Big],
	\label{H-odd}
	\end{align}
where $n\in\Z^+$, $\bsigma_\pm:= \text{\textbf{I}}\pm \bsigma_1$, $\bx_n:=(x_1,x_2,\cdots,x_n)$, and
	\begin{align}
	&c_{2n}(\bx_{2n}):=\prod_{j=1}^n\alpha(x_{2j-1})\nu(x_{2j}),
	&&c_{2n+1}(\bx_{2n+1}):=\alpha(x_1)\prod_{j=1}^n\nu(x_{2j})\alpha(x_{2j+1}),
	\label{cs-def}\\
	&d_{2n}(\bx_{2n}):=\prod_{j=1}^n\nu(x_{2j-1})\alpha(x_{2j}),
	&&d_{2n+1}(\bx_{2n+1}):=\nu(x_1)\prod_{j=1}^n\alpha(x_{2j})\nu(x_{2j+1}).
	\label{ds-def}
	\end{align} 

Next, we suppose that there are functions $w_\varepsilon:[0,1]\to\C$ and $w_\mu:[0,1]\to\C$ satisfying
	\begin{align}
	&\hat\varepsilon(x)=\left\{\begin{array}{ccc}
	w_{\varepsilon}(x/\ell)&\for&x\in[0,\ell],\\
	1&\for&x\notin[0,\ell],\end{array}\right.
	&&\hat\mu(x)=\left\{\begin{array}{ccc}
	w_{\mu}(x/\ell)&\for&x\in[0,\ell],\\
	1&\for&x\notin[0,\ell],\end{array}\right.
	\label{we-wm}
	\end{align}
and introduce  
	\begin{align}
	&\check c_1(\cx):=\check\alpha(\cx):=\left\{\begin{array}{ccc}
	w_\mu(\cx) &\for&\mbox{TE waves},\\
	w_\varepsilon(\cx)&\for&\mbox{TM waves},
	\end{array}\right.
	&&
	\check d_1(\cx):=\check\nu(\cx)
	:=\frac{w_\varepsilon(\cx)w_\mu(\cx)-\sin^2\theta}{\check\alpha(\cx)},	
	\label{ws-def}\\
	&\check c_{2n}(\check\bx_{2n}):=
	\prod_{j=1}^n\check\alpha(\cx_{2j-1}) \check\nu(\cx_{2j}),
	&&\check c_{2n+1}(\check\bx_{2n+1}):=
	\check\alpha(x_1)\prod_{j=1}^n\check\nu(\cx_{2j})\check\alpha(\cx_{2j+1}),
	\label{cns2}\\
	&\check d_{2n}(\check\bx_{2n}):=
	\prod_{j=1}^n\check\nu(\cx_{2j-1})\check\alpha(\cx_{2j}),
	&&\check d_{2n+1}(\check\bx_{2n+1}):=
	\check \nu(\cx_1)\prod_{j=1}^n\check\alpha(\cx_{2j})\check\nu(\cx_{2j+1}),
	\label{dns2}
	\end{align}
where $\cx:=x/\ell\in[0,1]$, $n\geq 1$, and for all $\bx_n\in[0,\ell]^n$,  $\check\bx_n$ stands for $\ell^{-1}\bx_n$, i.e., $\check\bx_n:=(\cx_1,\cx_2,\cdots,\cx_n)$. Then, for all $x\in[0,\ell]$, $n\geq 2$, and $\bx_n\in[0,\ell]^n$, we have
	\begin{align}
	&\hat\varepsilon(x)=w_\varepsilon(\check x),
	&&\hat\mu(x)=w_\mu(\check x),
	\label{ws}\\
	&\alpha(x)=\check\alpha(\cx)=\check c_1(\cx),
	&&\nu(x)=\check\nu(\cx)=\check d_1(\cx),
	\label{d1-t}\\
	&c_{n}( \bx_{n})=\check c_{n}(\check\bx_{n}),
	&&d_{n}( \bx_{n})=\check d_{n}(\check\bx_{n}).
	\label{c-d=}
	\end{align}
	
Equations (\ref{H-even}), (\ref{H-odd}), and \eqref{we-wm} -- \eqref{c-d=} allow us to simplify (\ref{dyson-M}) by changing the variables of integrations on its right-hand side from $x_j$'s to $\check x_j$'s. This gives 
	\begin{align}
	\bM = &\: e^{-ik\ell|\!\cos\theta|\bsigma_3}\bigg[\mathbf{I} + 
	\frac{1}{2} \sum_{n=1}^\infty \bigg\{
	\frac{(ik\ell)^{2n-1}}{|\!\cos\theta|} \left( {D}_{2n-1} \bcK+
	\cos^2\theta\,{C}_{2n-1} \bcK^T \right) + \nn\\
	&\hspace{4.2cm}
	(ik\ell)^{2n} \left(  {C}_{2n} \bsigma_-+{D}_{2n} \bsigma_+ \right) \bigg\}\bigg],
	\label{dyson-M2}
	\end{align}
where for all $n\geq 1$,
	\begin{align}
	&C_n:=\int_0^1d\cx_n\int_0^{\cx_{n}}d\cx_{n-1}\cdots\int_0^{\cx_2}d\cx_1\:
	\check c_n(\check\bx_n),
	\label{Cn=}\\
	&D_n:=\int_0^1d\cx_n\int_0^{\cx_{n}}d\cx_{n-1}\cdots\int_0^{\cx_2}d\cx_1\:
	\check d_n(\check\bx_n).
	\label{Dn=}
	\end{align}
In view of (\ref{bcK-def}), \eqref{Pauli}, and \eqref{dyson-M2}, the entries of the transfer matrix have the form
	\begin{align}
	M_{ab}=&\: e^{(-1)^{a} ik\ell |\!\cos\theta|}\bigg[\delta_{ab}+
	\sum_{n=1}^\infty S_{ab,n}(ik\ell)^n\bigg],
	\label{Mab=}
	\end{align}
where, for all $n\geq 1$,
	\begin{align}
	&\begin{aligned}
	&S_{ab,2n-1}:=\frac{(-1)^{a+1}}{2|\!\cos\theta|}\,\left[D_{2n-1}
	+(-1)^{a+b}\cos^2\theta\,C_{2n-1}\right],\\
	&S_{ab,2n}:=\frac{1}{2}\left[D_{2n}+(-1)^{a+b}C_{2n}\right].
	\end{aligned}
 	\label{Sab=}
 	\end{align}
	
We can substitute the Maclaurin expansion of the exponential factor on the right-hand side of \eqref{Mab=} and multiply it by the terms in the square bracket to obtain a series expansion of $M_{ab}$ in powers of $k\ell$. This gives the low-frequency expansions of the entries of the transfer matrix which we use in Sec.~\ref{s4} to obtain the low-frequency expansions of the reflection and transmission amplitudes. Truncating these series, we find low-frequency approximations whose accuracy depends on the convergence behavior of the series  $\sum_{n=1}^\infty S_{ab,n}(ik\ell)^n$. To study the latter, first we use (\ref{bounds}), \eqref{we-wm}, \eqref{ws-def} and \eqref{ws} -- \eqref{d1-t}, to show that, for all $\cx\in[0,1]$, $0<b_-\leq |\check c_1(\cx)|\leq b_+$ and $|\check d_1(\cx)|\leq(b_+^2+1)/b_-$. These inequalities suggest introducing 
	\[b_\star:= \frac{\sqrt{(b_+^2+1)^2+b_-^2b_+^2}}{b_-},\]
which satisfies $|\check c_1(\cx)|<b_\star$ and $|\check d_1(\cx)|<b_\star$. By virtue of these relations and  \eqref{cns2}, \eqref{dns2}, \eqref{Cn=}, and \eqref{Dn=}, we have
	\begin{align}
	&|\check c_n(\check\bx_n)|<b_\star^n, && |\check d_n(\check\bx_n)|<b_\star^n,
	&|C_n|< \frac{b_\star^n}{n!}, &&D_n< \frac{b_\star^n}{n!},
	\label{bound2}
	\end{align}
where $n\geq 1$ and $\bx_n\in[0,1]^n$. 
 
Because $|\!\cos\theta|<1$,  we can use \eqref{Sab=} and (\ref{bound2}) to establish		
	\begin{align}
	&|S_{ab,n}|(k\ell)^n<\frac{(|C_n|+|D_n|)(k\ell)^n}{2|\!\cos\theta|}
	<\frac{(b_\star k\ell)^n}{|\!\cos\theta| n!}.
	\label{bound3}
	\end{align}
This in turn implies that, for all $N\geq 1$,
	\[\sum_{n=1}^N |S_{ab,n}| (k\ell)^n< \frac{e^{b_\star k\ell}-1}{|\!\cos\theta|}.\]
Because $\cos\theta\neq 0$, this proves the absolute convergence of the series $\sum_{n=1}^\infty S_{ab,n}(ik\ell)^n$ for all values of $k\ell$. Therefore, according to (\ref{Mab=}), $M_{ab}$ are entire functions of $k\ell$. 
 
Furthermore, we can use \eqref{bound3} and the fact that $n!<\sqrt{2\pi n}(n/e)^n$ for $n\geq 2$ to conclude that the low-frequency approximation in which we neglect the contributions of the $n$-th and higher order terms in powers of $k\ell$ to $M_{ab}$ is valid for $k\ell\ll n (2\pi n\cos^2\theta)^{1/2n}/eb_\star$.
In terms of the incident wavelength, $\lambda:=2\pi/k$, this condition takes the form $\lambda\gg t_n b_\star\,\ell$,
	\be
	\lambda\gg t_n b_\star\,\ell,
	\label{LF-bound}
	\ee	
where
		\[t_n:=\frac{2 \pi e}{n\left(2\pi n \cos^2\theta\right)^{1/2n}}.\]
For $n=2$ and $3$, this gives $\lambda\gg 4.54\, b_\star\sqrt{|\sec\theta|}\ell$ and
$\lambda\gg 3.49\, b_\star|\sec\theta|^{1/3}\ell$, respectively. 

A useful consequence of \eqref{ws-def} -- \eqref{c-d=} and \eqref{Cn=} -- \eqref{Sab=} is that $M_{ab}$ admits a Laurent series expansion in powers of $|\!\cos\theta|$ of the form
	\be
	M_{ab}=\frac{M_{ab,-1}}{|\!\cos\theta|}+\sum_{m=0}^\infty M_{ab,m}|\!\cos\theta|^m,
	\label{grazing}
	\ee
where $M_{ab,m}$ are $\theta$-independent coefficients, i.e., $\theta=90^\circ$ marks a simple pole of $M_{ab}$. In Appendix~A we use (\ref{grazing}) to derive approximate expressions for the reflection and transmission amplitudes of our system at grazing angles where $|\!\cos\theta|\ll 1$.

We end this section by drawing attention to the fact that because $\det\bM=1$, the coefficients $C_n$ and $D_n$ entering the low-frequency expansion of the transfer matrix through \eqref{dyson-M2} are not independent. Using \eqref{Mab=} and \eqref{Sab=}, we have shown that $\det\bM=1$ holds for all $k\ell$ if and only if, for all $n\geq 1$,
	\be
	C_{2n}+D_{2n}=\sum_{m=1}^{2n-1}(-1)^{m-1}C_mD_{2n-m}.\nn
	\ee
For $n=1$, this gives the identity
	\be
	C_2+D_2=C_1D_1,
	\label{id1}
	\ee
which we can easily verify using \eqref{Cn=} and \eqref{Dn=}.

\section{Reflection and Transmission of Low-Frequency Waves}
\label{s4}

\subsection{{Low-frequency expansions of $R^{l/r}$ and $T$}}

Having expressed the entries of the transfer matrix as a series in nonnegative integer powers of $k\ell$, we can use \eqref{Rs=} and \eqref{T=} to obtain the low-frequency expansions of the reflection and transmission amplitudes. This requires inverting the series expansion of $M_{22}$ and multiplying the resulting series by the low-frequency series expansions of $M_{12}$ and $M_{22}$. In the following, we use this approach to compute the leading- and next-to-leading-order terms in the low-frequency expansions of the reflection and transmission amplitudes. 

First, we use \eqref{Sab=} and \eqref{id1} to express \eqref{Mab=} as
	 \begin{align}
	M_{ab}=&\:\delta_{ab}+ 
	\frac{i(-1)^{a+1}k\ell }{2|\!\cos\theta|} 
	\bigg[ D_1 +\cos^2\theta \Big\{ (-1)^{a+b} C_1-2\delta_{ab} \Big\} \bigg]+\nn\\
	&\frac{(k\ell)^2}{2} \bigg[ \Big\{1-(-1)^{a+b}\Big\}C_2-(C_1-1)D_1+	
	\cos^2\theta \Big\{(-1)^{a+b}C_1-\delta_{ab}\Big\} \bigg] + \cO(k\ell)^3,
	\label{Mab-O2}
	\end{align}
where $\cO(k\ell)^n$ stands for the sum of terms of order $n$ and higher in powers of $k\ell$. To make the $\theta$-dependence of $M_{ab}$ more transparent, we write $C_1$, $D_1$, and $C_2$ in the following more explicit form.
	\begin{align}
	&C_1=\int_{0}^{1}d\check{x}~\check \alpha({\check x}),
	\quad\quad\quad
	D_1=\int_{0}^{1}\! d\check{x}~\check \nu({\check x})
	=D_{1,0}+\cos^2\theta\,D_{1,1}, 
	\label{C1-D1}\\
	&C_{2}=
	\int_{0}^{1}\!\!d\check{x}_{2}\int_{0}^{\check{x}_2}\!\!d\check{x}_1\:
	\check\alpha(x_1)\check\nu(x_2)
	=C_{2,0}+\cos^2\theta\,C_{2,1},
	\label{C2=12}
	\end{align}
where we have benefitted from \eqref{alpha=}, \eqref{nu-def}, \eqref{ws} -- \eqref{c-d=}, \eqref{Cn=}, and \eqref{Dn=}, introduced
	\begin{align}
	&D_{1,0}:=\int_{0}^1 d\cx\left[\check\beta(\cx)-\frac{1}{\check\alpha(\cx)}\right],
	\quad\quad\quad
	~D_{1,1}:=\int_0^1 \frac{d\cx}{\check\alpha(\cx)},
	\label{D101-def}\\
	&C_{2,0}:=\int_{0}^{1}\!\!d{x}_2\int_{0}^{{\cx}_2}\!\!d{\cx}_{1}\:
	\check\alpha({\cx_1})\left[\check\beta(\cx_2)-\frac{1}{\check\alpha(\cx_2)}\right],\\
	&C_{2,1}:=\int_{0}^{1}\!\!d{\cx}_2\int_{0}^{{\cx}_2}\!\!d{\cx}_{1}\:
	\check\alpha({\cx_1})\check\alpha({\cx_2})^{-1},
	\label{C2-1-def}\\
	&\check\beta(\cx):=\beta(\ell\cx),
	\quad\quad\quad \beta(x):=\frac{\fn(x)^2}{\alpha(x)}=
	\left\{\begin{array}{lll}
	\hat{\varepsilon}(x) & \text { for } & \text {TE waves, } \\
	\hat{\mu} (x)& \text { for } & \text {TM waves},
	\end{array}\right.
	\label{beta-def}
	\end{align}
and used the fact that $\check\nu=\check\beta-\sin^2\theta/\check\alpha$.
	
Substituting \eqref{Mab-O2} in \eqref{Rs=} and \eqref{T=}, and making use of \eqref{D101-def}--\eqref{C2-1-def}, we find
	\begin{align}
	&R^{l/r}=R_1\,k\ell+R^{l/r}_2\,(k\ell)^2+\cO(k\ell)^3, 
	\label{R=2}\\
	&T=1+T_1\,k\ell+T_2\,(k\ell)^2+\cO(k\ell)^3, 
	\label{T=2}
	\end{align}
where
	\begin{align}
	&R_1:=R_{1,-1}|\sec\theta|+R_{1,1}|\!\cos\theta|,
	&&R^{l/r}_2:=R_{2,-2}\sec^2\theta+R^{l/r}_{2,0}+R^{l/r}_{2,2}\cos^2\theta,
	\label{rs=}\\
	&T_1:=T_{1,-1}|\sec\theta|+T_{1,1}|\!\cos\theta|,
	&&T_2:=T_{2,-2}\sec^2\theta+T_{2,0}+T_{2,2}\cos^2\theta,
	\label{ts=}
	\end{align}
and
	\begin{align}
	&R_{1,-1}:=T_{1,-1}=\tfrac{i}{2}D_{1,0},
	\label{R1m1-def}\\
	&R_{1,1}:=\tfrac{i}{2}(D_{1,1}-C_1),
	\label{R1p1-def}\\
	&R_{2,-2}:=T_{2,-2}:=-\tfrac{1}{4}D_{1,0}^2,
	\label{zR2m2}\\
	&R^{l}_{2,0}:=\tfrac{1}{2}[D_{1,0}(C_1-D_{1,1})-2C_{2,0}],\\
	&R^{l}_{2,2}:=\tfrac{1}{4}(C_1^2-D_{1,1}^2+2C_1D_{1,1}-4C_{2,1}),\\
	&R^{r}_{2,0}:=\tfrac{1}{2}[D_{1,0}(-C_1-D_{1,1}+2)+2C_{2,0}],\\
	&R^{r}_{2,2}:=\tfrac{1}{4}[C_1^2-D_{1,1}^2-2C_1D_{1,1}+4(-C_1+D_{1,1}+C_{2,1})],
	\label{zRr22}\\
	&T_{1,1}:=\tfrac{i}{2}(C_1+D_{1,1}-2),
	\label{zT11}\\
	&T_{2,0}:=-\tfrac{1}{2}D_{1,0}(D_{1,1}-1),\\
	&T_{2,2}:=-\tfrac{1}{4}\left[(C_1-1)^2+(D_{1,1}-1)^2\right].
	\label{t22=}
	\end{align}
For $\hat\varepsilon=\hat\mu=1$, we have $\check\alpha=\check\beta=1$, $C_1=D_{1,1}=2C_{2,1}=1$, and $D_{1,0}=C_{2,0}=0$. In view of (\ref{rs=}) -- (\ref{t22=}), these imply $R_1=R_2^{l/r}=T_1=T_2=0$, as expected. 

If the slab is made of nonmagnetic material and we are interested in the scattering of TE waves, $\alpha=\hat\mu=1$ and $\beta=\hat\varepsilon$. These imply $C_1=D_{1,1}=2C_{2,1}=1$ and
	\begin{align}
	&D_{1,0}=\int_0^1 d\cx[w_\varepsilon(\cx)-1]=\frac{1}{\ell}\int_0^\ell dx\:[\hat\varepsilon(x)-1],
	\label{C1-TE}\\
	&C_{2,0}=\int_0^1 d\cx \,\cx[w_{\varepsilon}(\cx)-1]=\frac{1}{\ell^2}\int_0^\ell dx\:x[\hat\varepsilon(x)-1],
	\label{C20-TE}\\
	&R^l=\frac{i}{2} D_{1,0} |\sec\theta|\,k\ell-
	\left[C_{2,0}+\frac{1}{4}D_{1,0}^2\sec^2\theta\right](k\ell)^2
	+\cO(k\ell)^3,
	\label{TE-RL}\\
	&R^r=\frac{i}{2} D_{1,0} |\sec\theta|\,k\ell+
	\left[C_{2,0}-\frac{1}{4}D_{1,0}^2\sec^2\theta\right](k\ell)^2+\cO(k\ell)^3,\\
	&T=1+\frac{i}{2}D_{1,0}|\sec\theta|\,k\ell-\tfrac{1}{4}D_{1,0}^2\sec^2\theta(k\ell)^2+
	\cO(k\ell)^3.
	\label{TE-T}
	\end{align}
For $\theta=0$, Eqs.~(\ref{TE-RL}) -- \eqref{TE-T} coincide with Eqs.~(52) -- (54) of Ref.~\cite{jpa-2021}, respectively. This provides a nontrivial check on the validity of our calculations.


\subsection{{Generalized Brewster's angle}}

{According to \eqref{rs=}, \eqref{R1m1-def}, and \eqref{R1p1-def}, the} leading order term $R_1$ in the low-frequency expansion of the $R^{l/r}$ vanishes provided that 
		\be
		|\!\cos\theta|=\sqrt{\frac{D_{1,0}}{C_1-D_{1,1}}}.
		\label{brw-1}
		\ee
In particular, $D_{1,0}(C_1-D_{1,1})^{-1}$ is a positive real number that does not exceed $1$. For a homogeneous slab where $\hat\varepsilon$ and $\hat\mu$ (and consequently $\alpha$ and $\beta$) take constant values inside the slab, $C_1=\alpha$, $D_{1,0}=\beta-\alpha^{-1}$, $D_{1,1}=\alpha^{-1}$, and (\ref{brw-1}) becomes
	\be
	|\!\cos\theta|=\sqrt{\frac{\alpha\beta-1}{\alpha^2-1}}=\sqrt{\frac{\fn^2-1}{\alpha^2-1}}.
	\label{brw-2}
	\ee	
For TM waves scattered by a nonmagnetic slab, $\alpha=\hat\varepsilon=\fn^2$, $\beta=\hat\mu=1$, and \eqref{brw-2} gives $|\!\cos\theta|
	=(\fn^2+1)^{-1/2}$, which is equivalent to $|\tan\theta|=|\fn|$. Therefore either $|\theta|$ or $|\theta-180^\circ|$ equals the Brewster's angle \cite{Born-Wolf}. This observation suggests writing (\ref{brw-1}) in the form
	\be
	\theta=\left\{
	\begin{array}{ccc}
	\pm\,\theta_{\rm GB}&\for&\mbox{left-incident waves},\\
	\pm\,\theta_{\rm GB}+180^\circ&\for&\mbox{right-incident waves},
	\end{array}\right.
	\label{gen-brw}
	\ee
where 
	\be
	\theta_{\rm GB}:={\rm arctan}\,\sqrt{\frac{C_1-D_{1,0}-D_{1,1}}{D_{1,0}}}=
	{\rm arctan}\,\sqrt{\frac{\int_0^\ell dx\,[\alpha(x)-\beta(x)]}{\int_0^\ell dx\,[\beta(x)-\alpha(x)^{-1}]}}.
	\label{GB-def}
	\ee
{This equation defines a generalization of the Brewster's angle;  for TM waves scattered by a homogeneous nonmagnetic slab, $\theta_{\rm GB}$ equals Brewster's angle, and for TE and TM waves scattered by an inhomogeneous slab, whenever the incidence angle is given by (\ref{gen-brw}), the linear terms in the low-frequency expansions of the reflection amplitudes vanish, i.e., $R^{l/r}$ become negligibly small at low frequencies. More precisely, (\ref{gen-brw}) implies}
	\begin{align}
	&R^{l/r}=R^{l/r}_2\,(k\ell)^2+\cO(k\ell)^3, &&
	R^l_2=-R^r_2=
	-C_{2,0}+\tfrac{1}{2}(C_1^2-2C_{2,1})\cos^2\theta_{\rm GB},
	\end{align}
and 
	\begin{align}
	T&=1+i(C_1-1)\cos\theta_{\rm GB}\;k\ell
	-\tfrac{1}{2}(C_1-1)^2\cos^2\theta_{\rm GB}(k\ell)^2+
	\cO(k\ell)^3\nn\\[3pt]
	&=e^{i(C_1-1)k\ell\cos\theta_{\rm GB}}+\cO(k\ell)^3.
	\end{align}
In particular, if $C_1=1$, which means $\frac{1}{\ell}\int_0^\ell dx\:\alpha(x)=1$,
we have $|T|=1+\cO(k\ell)^3$. This shows that if the incidence angle of a low-frequency TE or TM wave is given by (\ref{gen-brw}) and the average value of $\alpha$ over the slab is 1, the transmission of the wave through the slab does not change its amplitude.

\subsection{{Reciprocal reflection at low frequencies}}
	 
{According to \eqref{R=2}, $R^l-R^r=\cO(k\ell)^2$. This shows that} at low frequencies nonreciprocal reflection is a quadratic effect in the frequency of the wave. It is further suppressed, if $R_2^l=R_2^r$. {In view of \eqref{rs=} and \eqref{zR2m2} -- \eqref{zRr22}, this} is equivalent to
	\be
	\cos^2\theta=\frac{2C_{2,0}-D_{1,0}(C_1-1)}{C_1-2C_{2,1}+D_{1,1}(C_1-1)}.
	\label{condi2}
	\ee
If the right-hand side of this equation is a real number belonging to the interval $(0,1]$, there are two pairs of incidence angles at which $R^l-R^r=\cO(k\ell)^3$. Otherwise reciprocity in reflection is violated at all incidence angles whenever the quadratic terms in $k\ell$ are not negligibly small.

\subsection{{Transparency at low frequencies}}

{As seen from \eqref{ts=}, \eqref{R1m1-def}, and \eqref{zT11} -- \eqref{t22=},} transmission amplitude does not depend on $C_{2,0}$ and $C_{2,1}$. In particular, $T=1+\cO(k\ell)^3$ and the slab is transparent for low-frequency waves, if $C_1=D_{1,1}=1$ and $D_{1,0}=0$.\footnote{Under this condition, $R^l=-R_2(k\ell)^2+\cO(k\ell)^3$ and $R^r=R_2(k\ell)^2+\cO(k\ell)^3$, where $R_2:=C_{2,0}+\linebreak (C_{2,1}-\frac{1}{2})\cos^2\theta\, (k\ell)^2$.} This happens whenever
	\begin{align}
	&\int_0^1d\cx\:\check\alpha(\cx)=\int_0^1 d\cx\:\check\beta(\cx)=
	\int_0^1  \frac{d\cx}{\check\alpha(\cx)}=1,\nn
	\end{align}
or equivalently
	\begin{align}
	&\frac{1}{\ell}\int_0^\ell dx\:\hat\varepsilon(x)=\frac{1}{\ell}\int_0^\ell dx\:\hat\mu(x)=1~~~{\rm and}~~~\frac{1}{\ell}\int_0^\ell  \frac{dx}{\alpha(x)}=1.
	\label{transparent}
	\end{align}
This condition means that the average values of $\hat\varepsilon$, $\hat\mu$, and $\alpha^{-1}$ over the slab must be equal to $1$. This is the case, if $\hat\varepsilon(x)=1+\fz_1\,f_1(x/\ell)$ and $\hat\mu(x)=1+\fz_2\,f_2(x/\ell)$, where $\fz_1$ and $\fz_2$ are real or complex coefficients such that $|\fz_1|\ll 1$ and $|\fz_2|\ll 1$, and $f_1$ and $f_2$ are bounded functions such that $f_1(\cx)=f_2(\cx)=0$ for $\cx\notin[0,1]$ and $\int_0^1d\cx\,f_1(\cx)=\int_0^1d\cx\,f_1(\cx)=0$. A simple example is $f_{1}(\cx):=e^{2i\pi n\cx}$ and $f_{2}(\cx):=0$, where $n$ is an integer. This is the principal example of a permittivity profile displaying perturbative unidirectional invisibility \cite{lin-2011,longhi-2011,pra-2015b}. In particular, it is transparent for sufficiently small values of $|\fz_1|$.

\subsection{{Implications of $\cP\cT$-symmetry}}

Consider the case where the slab is $\cP\cT$-symmetric, i.e., $\hat\varepsilon(\ell-x)^*=\hat\varepsilon(x)$ and $\hat\mu(\ell-x)^*=\hat\mu(x)$. Then under the parity transformation (space reflection), $x\to\ell-x$, the real parts of $\hat\varepsilon$ and $\hat\mu$ are left invariant (are even), and their imaginary parts change sign (are odd).\footnote{We call a function $f:\R\to\C$ even (respectively odd) with respect to the parity transformation $x\to\ell-x$, if $f(\ell-x)=f(x)$ (respectively $f(\ell-x)=-f(x)$).} This implies that the same applies to $\alpha$ and $\beta$. Therefore, $|\alpha|$ is even, and because $\alpha^{-1}=|\alpha|^{-2}\alpha^*$, the real and imaginary parts of $\alpha^{-1}$ are also respectively even and odd. In light of the transformation properties of $\alpha$, $\beta$, and $\alpha^{-1}$ under $x\to\ell-x$ and Eqs.~\eqref{C1-D1} and \eqref{D101-def}, 
	\begin{align}
	&C_1=\int_{0}^{1}d\check{x}~\RE[\check \alpha({\check x})],
	&&D_{1,0}=\int_{0}^1 d\cx\:\RE\!\left[\check\beta(\cx)-\frac{1}{\check\alpha(\cx)}\right],
	&&D_{1,1}=\int_0^1d\cx\:\frac{\RE[\alpha(\cx)]}{|\check\alpha(\cx)|^2},
	\label{CDD=}
	\end{align}
where ``$\RE$'' stands for the real part of its argument. Eqs.~\eqref{CDD=} allow us to express the low-frequency transparency condition \eqref{transparent} for $\cP\cT$-symmetric slabs in the form
	\begin{align}
	&\frac{1}{\ell}\int_0^\ell dx\:\RE[\hat\varepsilon(x)]=\frac{1}{\ell}\int_0^\ell dx\:\RE[\hat\mu(x)]=1~~~{\rm and}~~~\frac{1}{\ell}\int_0^\ell  dx\: \RE[\alpha(x)^{-1}]=1.
	\label{transparent-PT}
	\end{align}
Notice that (\ref{transparent-PT}) coincides with the transparency condition~(\ref{transparent}) for a slab with no gain or loss.

\subsection{{Generalization to $a\neq 0$}}

The formulas for the low-frequency expansions of the entries of the transfer matrix and the reflection and transmission amplitudes that we have given above apply to the cases where $a=0$. We can generalize them to situations where $a\neq 0$ using transformation property of the transfer matrix under the translation $x\to\tilde x:=x-a$. 

Let $\sS$ be the medium with inhomogeneities confined to a slab lying in the region given by $a\leq x\leq a+\ell$, and $\tilde\sS$ be the translated medium whose relative permittivity and relative permittivity are respectively given by $\hat{\tilde\varepsilon}(x):=\hat\varepsilon(x+a)$ and $\hat{\tilde\mu}(x):=\hat\mu(x+a)$. Denoting the transfer matrices of $\sS$ and $\tilde\sS$ respectively by $\bM$ and $\tilde\bM$, and recalling that (\ref{Mab=}) applies to $\tilde\sS$, we can express the entries of $\tilde\bM$ as
	\begin{align}
	\tilde M_{ab}=&\: e^{(-1)^{a} ik\ell |\!\cos\theta|}\bigg[\delta_{ab}+
	\sum_{n=1}^\infty S_{ab,n}(ik\ell)^n\bigg].
	\label{tMab=}
	\end{align}
Here $S_{ab,n}$ are given by \eqref{Sab=} and the coefficients $C_n$ and $D_n$ appearing in this equation are to be computed after we make the following transformations in \eqref{we-wm}.
	\begin{align}
	&\hat\varepsilon(x)\to\hat{\tilde\varepsilon}(x):=\hat\varepsilon(x+a),
	&&\hat\mu(x)\to\hat{\tilde\mu}(x):=\hat\mu(x+a).
	\label{trans-3}
	\end{align}
This identifies $w_\varepsilon$ and $w_\mu$ with the functions satisfying
	\begin{align}
	&\hat\varepsilon(x)=\left\{\begin{array}{ccc}
	w_{\varepsilon}\big((x-a)/\ell\big)&\for&x\in[a,a+\ell],\\
	1&\for&x\notin[a,a+\ell],\end{array}\right.\nn\\
	&\hat\mu(x)=\left\{\begin{array}{ccc}
	w_{\mu}\big((x-a)/\ell\big)&\for&x\in[a,a+\ell],\\
	1&\for&x\notin[a,a+\ell].\end{array}\right.
	\nn
	\end{align}	
Therefore, (\ref{trans-3}) amounts to redefining $\cx$ in \eqref{ws-def} -- \eqref{dns2} as $\cx:=(x-a)/\ell$.  

With the help of Eqs.~\eqref{asympsolns} and \eqref{M-def} written for $\sS$ and $\tilde\sS$, we can show that $\tilde{\bM}= e^{-i\mathfrak{K}a \bsigma_3}\,\bM \,e^{i\mathfrak{K}a \bsigma_3}$, \cite{tjp-2020}. Expressing this relation in terms of the entries of $\bM $ and $\tilde\bM $, we have
	\begin{align}
	&{M}_{11}= \tilde M_{11} , && {M}_{12} = e^{2i\fK a} \tilde M_{12},
	&& {M}_{21}= e^{-2i\fK a} \tilde M_{21}, &&  {M}_{22} =\tilde M_{22}.
	\label{M-trans}
	\end{align}
These equations together with (\ref{Rs=}) and (\ref{T=}) allow us to relate the reflection and transmission amplitudes of $\sS$ to those of $\tilde\sS$. Using a tilde to label the latter, we have
	\begin{align}
	&R^l =e^{-2i\fK a}\tilde R^l,
	&&R^r =e^{2i\fK a}\tilde R^r,
	&&T=\tilde T.
	\nn	
	\end{align}
Noting that it is $\tilde R^{l/r}$ and $\tilde T$ that are given by the right-hand sides of (\ref{R=2}) and \eqref{T=2}, we see that \eqref{T=2} also applies to cases where $a\neq 0$, and that the low-frequency expansions of the reflection amplitudes for these cases have the form
	\begin{align}
	&\begin{aligned}
	&R^l= e^{-2ika\cos\theta}[R_1k\ell+R^l_2(k\ell)^2]+\cO(k\ell)^3,\\
	&R^r= e^{2ika\cos\theta}[R_1k\ell+R^r_2(k\ell)^2]+\cO(k\ell)^3,
	\end{aligned}
	\label{RRT-general}
	\end{align}
where $R_1$ and $R^{l/r}_2$ are given by \eqref{rs=}, and coefficients entering the right-hand sides of \eqref{R1m1-def} -- \eqref{t22=} are to be computed after performing the transformation \eqref{trans-3}.

To provide a check on the validity of (\ref{RRT-general}) we applied these formulas to the scattering of TE and TM waves by a homogeneous slab whose reflection and transmission amplitudes admit exact analytic expressions. In Appendix~B, we compute $R_1$, $R_2^{l/r}$, $T_1$, and $T_2$ for this slab. Substituting the resulting expressions in (\ref{R=2}) and (\ref{T=2}), we find the very same expressions as the ones we obtain by expanding the right-hand sides of \eqref{RR-homogen} and \eqref{T-homogen} in powers of $k\ell$ and keeping the linear and quadratic terms.

\begin{figure}
	\begin{center}
        \includegraphics[scale=.65]{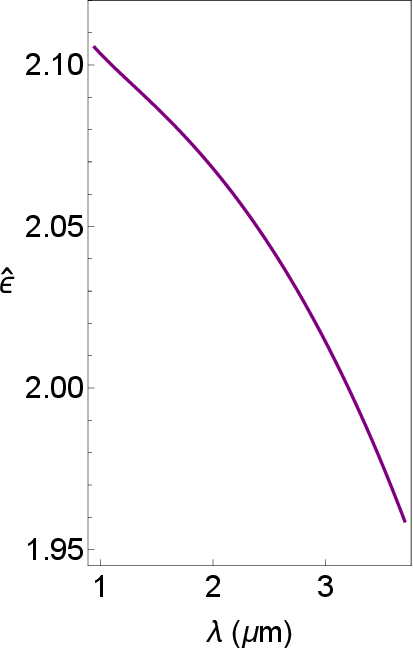}~
         \includegraphics[scale=.44]{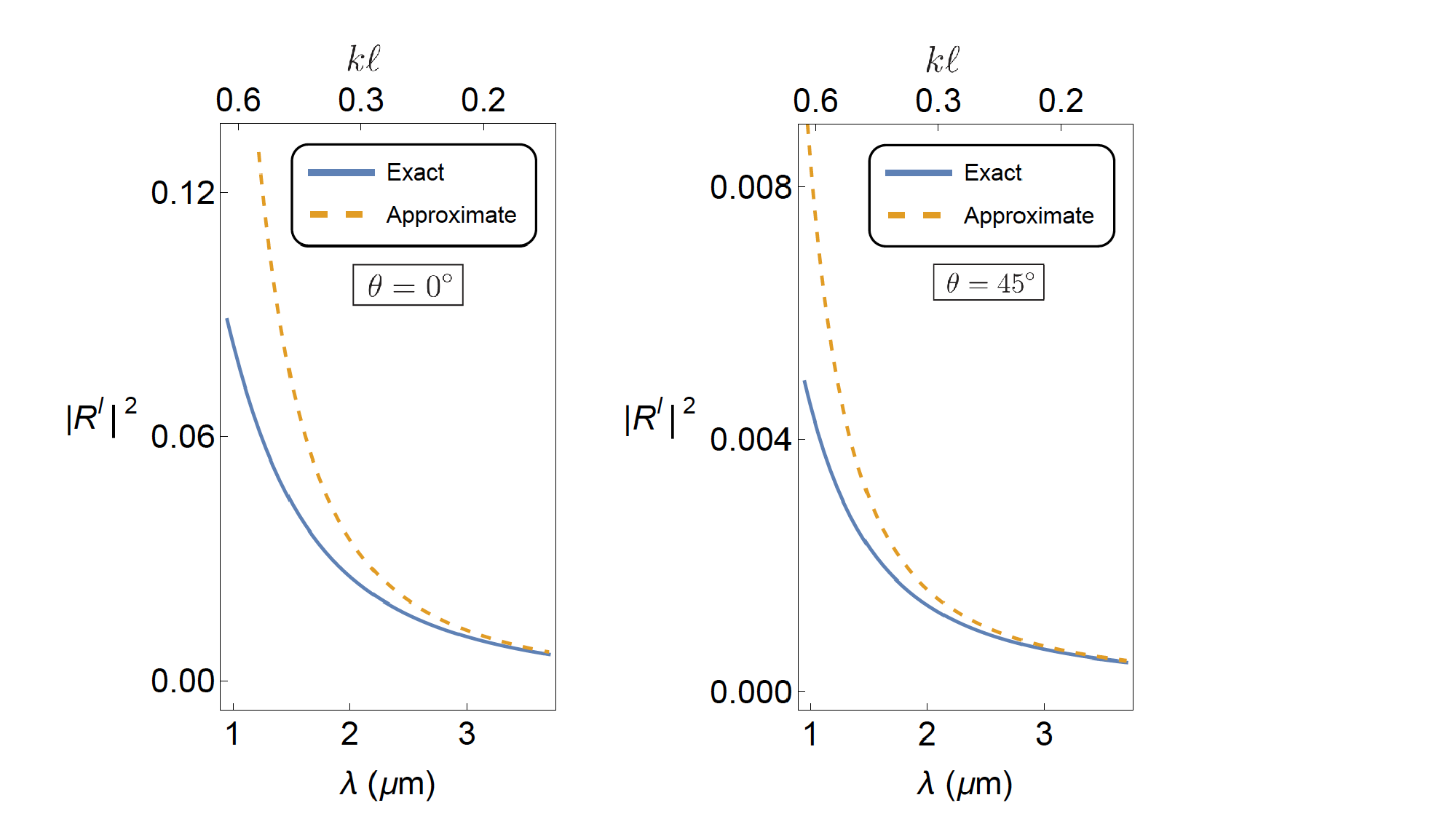}~
       \includegraphics[scale=.44]{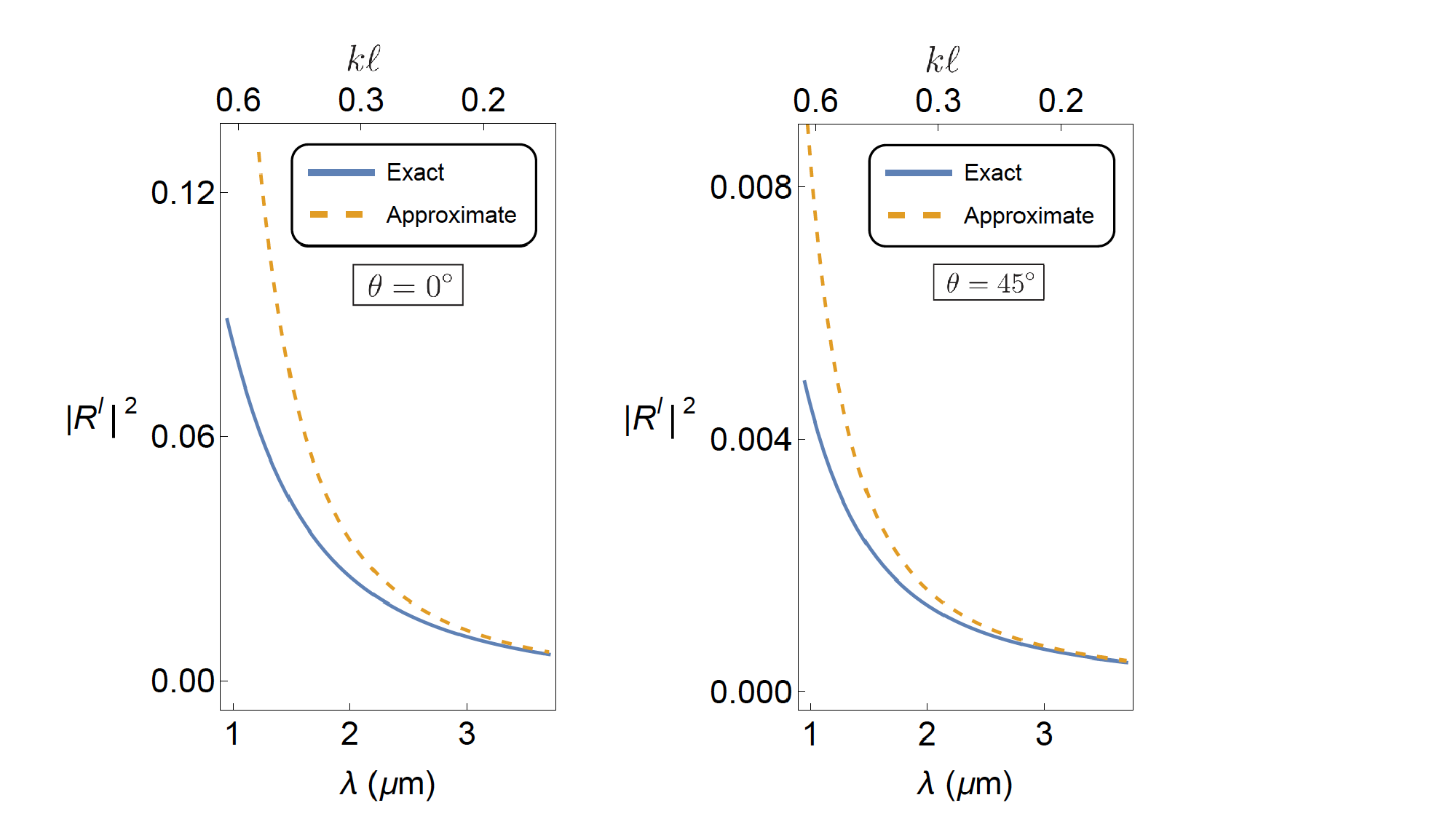}        
	\caption{Plots of the relative permittivity $\hat\varepsilon$ of a slab of thickness 0.1 $\mu$m made of fused Silica {as a function of the wavelength $\lambda$ and plots of its reflection coefficient $|R^{l}|^2$ as functions of $\lambda$ and $k\ell$} for incidence angles $\theta=0$ and $\theta=45^\circ$. The solid curves in the middle and right panels correspond to the exact values of $|R^{l}|^2$ given by \eqref{RR-homogen} while the dashed curves represent the outcome of the second-order low-frequency approximation. For $\lambda \gtrapprox3~\mu{\rm m}$ which corresponds to $k\ell\lessapprox0.21$, they agree with the exact results.}
        \label{fig2}
        \end{center}
        \end{figure}

To examine the domain of validity of the second-order low-frequency approximation, in which we neglect the cubic and higher order terms in the low-frequency expansions of the scattering data, we provide in Fig.~\ref{fig2} the plots of the exact and approximate expressions for the reflection coefficient $|R^{l}|^2$ associated with the scattering of TM waves by a homogeneous slab made of fused Silica (SiO$_2$). In this case $\beta=\hat\mu=1$ and $\alpha=\hat\varepsilon$. To generate these plots, we have taken $\ell=0.1~\mu{\rm m}$ and employed the following dispersion relation which is valid for wavelengths $\lambda$ in the range $0.21$ -- $3.71\,\mu{\rm m}$, \cite{Malitson-1965}.
	\be
	\hat\varepsilon=1+\frac{0.696\lambda^2}{\lambda^2-(0.0684\, \mu{\rm m})^2}+
	\frac{0.408\lambda^2}{\lambda^2-(0.116\, \mu{\rm m})^2}+\frac{0.897\lambda^2}{\lambda^2-(9.896\, \mu{\rm m})^2}.
	\nn
	\ee
For these wavelengths the graphs of $|R^l|^2$ have similar behavior for different values of the incidence angle $\theta$ away from $90^\circ$. For this reason, we only plot $|R^{l}|^2$ for $\theta=0^\circ$ and $\theta=45^\circ$. According to \eqref{LF-bound}, for these incidence angles, the second-order low-frequency approximation should be valid for
	\[\lambda\gg\left\{\begin{array}{ccc}
	2.10\, \mu{\rm m}&\for&\theta=0^\circ,\\
	2.36\, \mu{\rm m}&\for&\theta=45^\circ.\end{array}\right.\]
Fig.~\ref{fig2} shows that this approximation is actually reliable for $\lambda \gtrapprox 3\,\mu{\rm m}$.

\section{Low-Frequency Scattering in Half-Space}
\label{s5}

\begin{figure}
	\begin{center}
        \includegraphics[scale=.35]{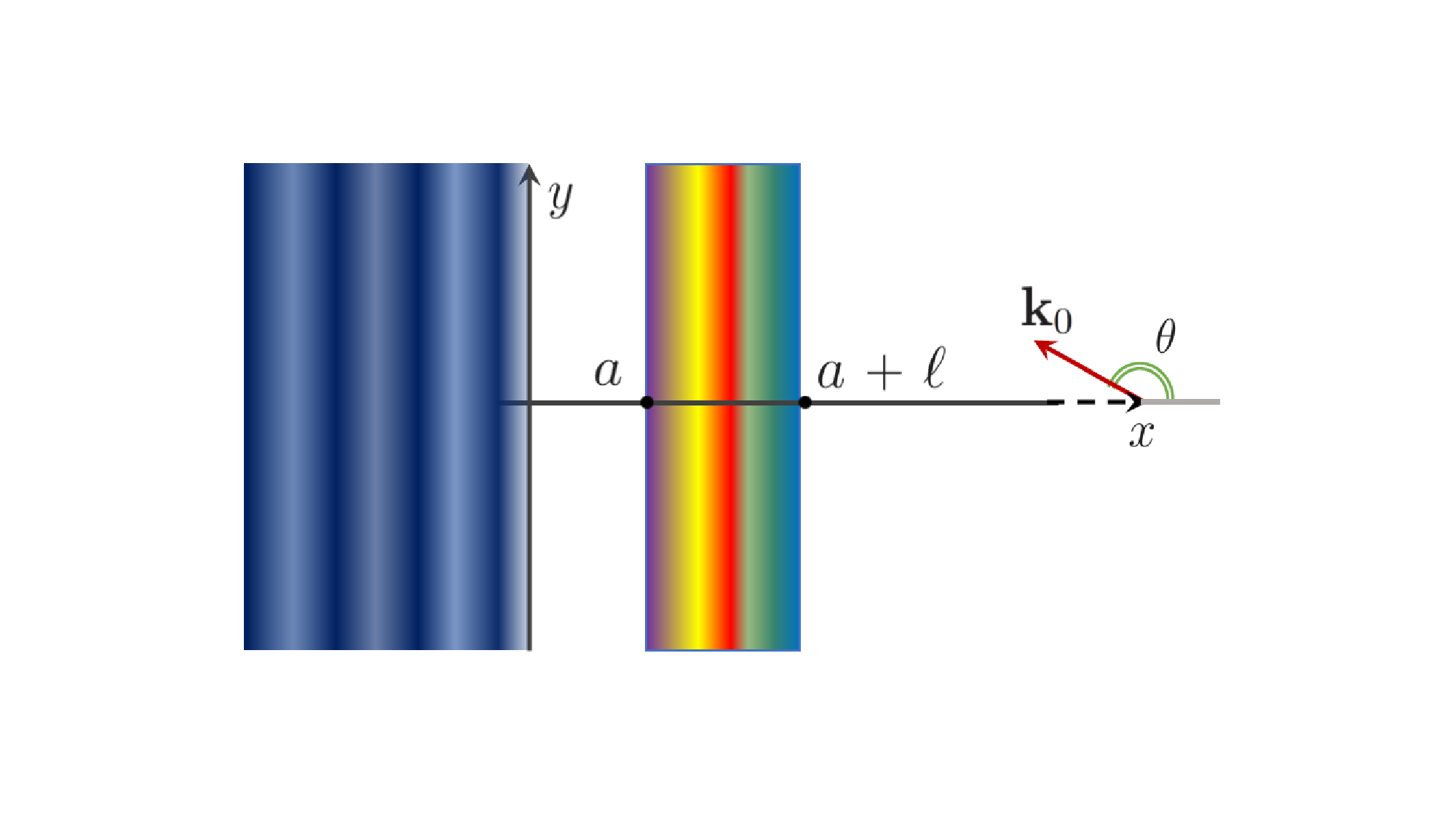}\vspace{-1cm}
        \caption{Schematic view of an inhomogeneous slab placed in the half-space $S^+$ given by $x>0$. The region painted in different shades of blue represents the material filling the half-space $S^-$. The source of the incident wave is placed on the plane $x=+\infty$. The incident wave vector $\bk_0$ is shown as a red arrow. The incidence angle $\theta$ ranges over the interval $(90^\circ,270^\circ)$.}
        \label{fig3}
        \end{center}
        \end{figure}

Suppose that our slab is located in the half-space $S^+$ defined by $x>0$, i.e., it occupies the region given by $a\leq x\leq a+\ell$ for some $a>0$, and the reset of the space ($S^-$) is filled with a material exhibiting planar symmetry, as shown in Fig.~\ref{fig3}. 
Then the effect of the content of $S^-$ on the waves propagating in $S^+$ can be encoded into a boundary condition on the interface, namely the $y$-$z$ plane. Assuming that this is a boundary condition of Robin type, 
	\begin{equation}
	\xi\,\phi(0) + \zeta\,\fK^{-1}\partial_x \phi(0) = 0, 
	\label{Robin}
	\end{equation}
where $\xi$ and $\zeta$ are complex coefficients possibly depending on $k$ and not vanishing simultaneously, i.e., $|\xi|+|\zeta|>0$, we can describe the TE and TM waves propagating in $S^+$ using the solutions of the Bergmann's equation (\ref{Bergmann}) in the positive half-line $\R^+$ that fulfill \eqref{Robin}.\footnote{Notice that the multiplication of $\xi$ and $\zeta$ by an arbitrary nonzero complex coefficient does not affect (\ref{Robin}).}

For $\xi\neq 0=\zeta$ and $\zeta\neq 0=\xi$, \eqref{Robin} reduces to the Dirichlet and Neumann boundary conditions, $\phi(0)=0$ and $\partial_x\phi(0)=0$, respectively. The first of these corresponds to placing the slab at a distance $a$ from a perfect mirror. 

Because the inhomogeneities of the medium $\sS$ are confined to the slab, solutions of (\ref{Bergmann}) satisfy
	\begin{equation}
	\phi(x) = N\left[e^{-i \fK x} + \cR\, e^{i \fK x}\right]~~\for~~ x \geq a+\ell,
	\label{half-line-asymp}
	\end{equation}
where $N$ stands for the amplitude of the incident wave, and $\cR$ denotes the complex reflection amplitude of the slab. The first term on the right-hand side of \eqref{half-line-asymp} represents an incident wave {with incidence angle $\theta$ satisfying $90^\circ<\theta<270^\circ$}. The quantity defined by $\cA:=1-|\cR|^2$ is called the absorption coefficient of the material filling $S^-$.

Ref.~\cite{ap-2019} shows that the potential scattering problems defined on a half-line can be reduced to the determination of the transfer matrix for certain potential scattering problems defined on the full line. The same construction applies to the scattering phenomena defined by the Bergmann's equation (\ref{Bergmann}) on the half line, because it relies on the relations between the entries of the transfer matrix and the reflection and transmission amplitudes which are identical to the ones for potential scattering. Combining the construction given in Ref.~\cite{ap-2019} and the results of Sec.~\ref{s3} of the present article, we can obtain the low-frequency expansions of the reflection amplitude $\cR$ and the absorption coefficient $\cA$.

The main result of Ref.~\cite{ap-2019} (its Eq.~(20)) is the following expression for the reflection amplitude.
	\be
	\cR=\frac{M_{11}-\gamma\, M_{12}}{M_{21}-\gamma\,M_{22}},
	\label{CR1}
	\ee
where $M_{ab}$ are the entries of the transfer matrix for the slab in the absence of the material filling $S^-$, i.e., when $\hat\varepsilon(x)=\hat\mu(x)=1$ for $x<0$, and
	\begin{equation}
	\gamma := \frac{\xi+ i\zeta}{\xi - i\zeta}.
	\label{gamma-def}
	\end{equation}
If the material filling the half-space $S^-$ is absent, $\phi(x)=T e^{-i\fK x}$ for $x\leq 0$, and we have $\phi(0)-i\fK^{-1}\partial_x\phi(0)=0$. We can express this condition in the form (\ref{Robin}), if we set $\xi=i\zeta\neq 0$. Because $|\xi|+|\zeta|>0$, this hold if and only if $\gamma=\infty$. In other words, in the presence of an inhomogeneous material filling $S^{-}$, $\gamma$ takes a finite value. In particular, Dirichlet and Neumann boundary conditions correspond to $\gamma=1$ and $-1$, respectively.
	 	
Eq.~\eqref{CR1} allows us to use the results of Sec.~\ref{s3} to determine  the low-frequency expansion of the reflection amplitude $\cR$. To do this, first we make use of \eqref{M-trans} to express \eqref{CR1} as 
	\begin{equation}
	\cR=\frac{e^{2i\fK a}[\tilde M_{11}-\gamma\, e^{2i\fK a} \tilde M_{12}]}{\tilde M_{21} - \gamma\, e^{2i\fK a}\tilde M_{22}}=
	\frac{\eta[\tilde M_{11}-\eta\,\tilde M_{12}]}{\gamma[\tilde M_{21} - \eta\,\tilde M_{22}]},
	\label{R-hspace}
	\end{equation}
where 
	\be
	\eta:=\gamma\, e^{2i\fK a}=\gamma\, e^{2ika|\!\cos\theta|}.
	\label{eta-def}
	\ee
Substituting \eqref{tMab=} in \eqref{R-hspace} and expanding the resulting expression as a power series in $k\ell$, we find
	\be
	\cR=-\gamma^{-1}+\cR_1 k\ell+\cR_2(k\ell)^2+\cO(k\ell)^3,
	\label{cR=2}
	\ee
where 
	\begin{align}
	&\cR_1:=\cR_{1,-1}|\sec\theta|+\cR_{1,1}|\!\cos\theta|,
	\label{cR1=}\\
	&\cR_2:=\cR_{2,-2}\sec^2\theta+\cR_{2,0}+\cR_{2,2}\cos^2\theta,\\
	&\cR_{1,-1}:= \frac{i\eta_-^2 D_{1,0}}{2\gamma\,\eta},
	\quad\quad\quad
	\cR_{1,1}:= \frac{i(\eta_-^2 D_{1,1}-\eta_+^2 C_1+4\eta)}{2\gamma\,\eta},
	\label{cR112=}\\
	&\cR_{2,0}:= -\frac{\eta_-}{2\gamma\,\eta^2}\Big[
	(\eta^2-1)C_1D_{1,0}+\eta_-D_{1,0}(\eta_-D_{1,1}-2\eta)
	-2\eta\,\eta_+C_{2,0}	\Big] ,
	\label{cR20=}\\
	&\cR_{2,-2}:=-\frac{\eta_-^3D_{1,0}^2}{4\gamma\,\eta^2},\\
	&\begin{aligned}
	\cR_{2,2}:=&\frac{1}{4\gamma\,\eta^2}\Big[
	\eta_+^3C_1^2-\eta_-^3D_{1,1}^2
	-2\eta_+C_1(\eta_-^2D_{1,1}+2\eta\eta_+)+\\
	&\hspace{1.3cm}
	4\eta[(\eta^2-1)C_{2,1}+2\eta]+4\eta\eta_-^2D_{1,1}\Big]
	\end{aligned}
	\label{cR22=}\\
	&\eta_\pm:=\eta\pm 1=\gamma\, e^{2ika|\!\cos\theta|}\pm 1.
	\label{eta-pm}
	\end{align}
	
The following are consequences of \eqref{cR=2} -- \eqref{eta-pm}.
	\begin{enumerate}
	\item For cases where $\sS$ is a nonmagnetic medium and we are interested in the scattering of TE waves, \eqref{C1-D1}, \eqref{D101-def}, and  \eqref{C2-1-def} give $C_1=D_{1,1}=2C_{2,1}=1$. Substituting these in \eqref{cR112=}, \eqref{cR20=}, and \eqref{cR22=}, we find $\cR_{1,1}=\cR_{2,2}=0$ and $\cR_{2,0}=(\eta^2-1)C_{2,0}/\gamma\eta$, and (\ref{cR=2}) becomes
	\be
	\cR=-\frac{1}{\gamma}\left[1-\frac{i\eta_-^2 D_{1,0} |\sec\theta|k\ell}{2\eta} -\frac{\eta_-(4\eta\eta_+C_{2,0}-\eta_-^2D_{1,0}^2\sec^2\theta)
	(k\ell)^2}{4\eta^2}\right]+\cO(k\ell)^3,
	\label{cR-TE=}
	\ee
where, in view of transformation \eqref{trans-3},
	\begin{align}
	&D_{1,0}=\frac{1}{\ell}\int_0^{\ell} d\tilde x\,[\hat{\tilde\varepsilon}(\tilde x)-1]=
	\frac{1}{\ell}\int_a^{a+\ell} dx\,[\hat\varepsilon(x)-1],\\
	&C_{2,0}=\frac{1}{\ell}\int_0^{\ell} d\tilde x\,\tilde x[\hat{\tilde\varepsilon}(\tilde x)-1]=\frac{1}{\ell^2}\int_a^{a+\ell} dx\:(x-a)[\hat\varepsilon(x)-1].
	\end{align}
For a normally incident TE wave, $\theta=180^\circ$, and (\ref{cR-TE=}) reduces to Eq.~(61) of Ref.~\cite{jpa-2021}.

\item Suppose that $\gamma=1$, which corresponds to a slab placed in front of a perfect mirror, and $k=\pi n/a|\!\cos\theta|$ for some $n\in\Z^+$. Then according to \eqref{eta-def} and (\ref{cR-TE=}), $\eta_-=0$ and  $\cR=-1+\cO(k\ell)^3$. This shows that we can make a nonmagnetic slab placed in front of a perfect mirror effectively reflectionless for low-frequency TE waves with incidence angle $\theta$ and wavelength $\lambda$ provided that its distance to the mirror is given by
	\be
	a=\frac{n\lambda}{2|\!\cos\theta|}
	\label{condi-103}
	\ee
for a positive integer $n$. If the slab is made of a magnetic material,  Condition \eqref{condi-103}, which is equivalent to $\gamma=\eta=1$, implies $\cR_{1,-1}=\cR_{2,0}=\cR_{2,-2}=0$, $\cR_{1,1}=2i(1-C_1)$, $\cR_{2,2}=2(C_1-1)^2$, and consequently
	\begin{align}
	\cR&=-1-2i(C_1-1)|\!\cos\theta|k\ell+2(C_1-1)^2\cos^2\theta\,(k\ell)^2+\cO(k\ell)^3
	\nn\\
	&=-e^{2i(C_1-1)|\!\cos\theta|\,k\ell}+\cO(k\ell)^3.\nn
	\end{align}
According to this equation, if $C_1$ happens to be a real number, $|\cR|^2=1+\cO(k\ell)^3$ and $\cA=\cO(k\ell)^3$. Therefore, the slab does not absorb low-frequency TE and TM waves. This is the case for a $\cP\cT$-symmetric slab, because in view of \eqref{C1-D1} for such a slab the imaginary part of $\int_{a}^{a+\ell}dx~\alpha(x)$ vanishes. If $C_1=1$, which means 
	$\frac{1}{\ell}\int_{a}^{a+\ell}dx~\alpha(x)=1$,
we have $\cR=-1+\cO(k\ell)^3$, i.e., the slab is reflectionless for low-frequency waves. 

\end{enumerate}

\section{Summary and Concluding Remarks}
\label{s6}

The scattering problem for TE and TM waves propagating in an effectively one-dimensional isotropic medium can be reduced to the determination of an associated transfer matrix $\bM$. The latter shares the basic properties of the standard transfer matrix of potential scattering in one dimension. In particular, it can be expressed in terms of the evolution operator for a non-unitary two-level quantum system. This gives a Dyson series expansion for $\bM$. For the scattering setups in which the scattering arises due to the inhomogeneities of the medium confined to a planar slab, we have used this observation to devise a systematic method for computing the coefficients of the low-frequency expansions of the transfer matrix and the reflection and transmission amplitudes of the medium. 

We have demonstrated the utility of this method in deriving explicit analytic formulas for the leading- and next-to-leading-order terms in the low-frequency expansions of the reflection and transmission amplitudes. These allowed us to identify a generalization of Brewster angle for low-frequency wave scattering by an inhomogeneous slab. They also revealed explicit conditions for the transparency and reflectionlessness of $\cP\cT$-symmetries and non-$\cP\cT$-symmetric  slabs at low frequencies.  

Our approach to low-frequency scattering can also be applied to situations where the scattering problem is defined in a half-line. This corresponds to TE and TM waves propagating in a half-space while the other half-space is filled with a material whose effect on the wave is determined in terms of a boundary condition of the Robin type at the interface. For this setup, we obtain analytic formulas for the low-frequency expansion of the reflection amplitude and discuss some of their concrete implications. 

Our results can be directly employed in the study the scattering of low-frequency acoustic waves propagating in a compressible fluid with planar symmetry. This is  because these waves are also described by Bergmann's equation in one dimension.  

For TE waves scattered by a non-magnetic slab with translational symmetry along a single direction, Maxwell's equations reduce to the Helmholtz equation in two-dimensions which is equivalent to time-independent Schr\"odinger equation in two-dimensions. There is a fundamental transfer matrix formulation of the scattering problem defined by the latter \cite{pra-2021} which also provides an effective method of computing the low-frequency expansions of the scattering data \cite{pra-2025}. The extension of this approach to TM waves and magnetic material requires a generalization of the notion of fundamental transfer matrix to the scattering problems defined by the Bergmann's equation in two dimensions. This is a nontrivial open problem which we hope to address in a future publication.

\subsection*{Acknowledgements} 
This work has been supported by the Scientific and Technological Research Council of T\"urkiye (T\"UB\.{I}TAK) in the framework of the project 123F180 and by Turkish Academy of Sciences (T\"UBA).

\section*{Appendix A: Scattering at grazing angles}

According to \eqref{ws-def} -- \eqref{c-d=} and \eqref{Cn=} -- \eqref{Sab=}, the coefficient of the principal part of the Laurent series (\ref{grazing}) has the form 
	\be
	M_{ab,-1}:=\frac{(-1)^{a+1}}{2}	\sum_{n=1}^\infty D_{2n-1,0}(ik\ell)^{2n-1},
	\label{grazing-2}
	\ee
where 
	\begin{align}
	&D_{2n-1,0}:=D_{2n-1}\Big|_{\theta=90^\circ}=
	\int_0^1d\cx_{2n-1}\int_0^{\cx_{2n-2}}d\cx_{2n-1}\cdots\int_0^{\cx_2}d\cx_1\:
	\check d_{2n-1,0}(\check\bx_{2n-1}),\nn\\
	&\check d_{1,0}(\cx):=\check d_1(\cx)\Big|_{\theta=90^\circ}=
	\check\nu(\cx)\Big|_{\theta=90^\circ}=\check\beta(\cx)-\check\alpha(\cx)^{-1},\nn\\
	&\check d_{2n+1,0}(\check\bx_{2n+1}):=\check d_{2n+1}(\check\bx_{2n+1})\Big|_{\theta=90^\circ}=
	[\check\beta(\cx_1)-\check\alpha(\cx_1)^{-1}]\prod_{j=1}^n
	\check\alpha(\cx_{2j})[\check\beta(\cx_{2j+1})-\check\alpha(\cx_{2j+1})^{-1}],\nn
	\end{align}
and we have employed \eqref{beta-def}.

For grazing angle(s) where $|\!\cos\theta|\ll 1$, \eqref{grazing} and \eqref{grazing-2} imply
	\be
	M_{ab}=\frac{(-1)^{a+1}}{2|\!\cos\theta|}\left[
	\sum_{n=1}^\infty D_{2n-1,0}(ik\ell)^{2n-1}+
	\cO(|\!\cos\theta|)\right].
	\label{theta-exp}
	\ee
Substituting (\ref{theta-exp}) in (\ref{Rs=}) and (\ref{T=}) and considering the generic cases where $M_{22,-1}\neq 0$, we have
	\begin{align}
	&R^{l/r}=-1+\cO(|\!\cos\theta|),
	&&
	T=\frac{-2|\!\cos\theta|}{\sum_{n=1}^\infty D_{2n-1,0}(ik\ell)^{2n-1}}+\cO(|\!\cos\theta|)^2.
	\end{align}

\section*{Appendix B: Low-frequency scattering by a homogeneous slab}
For a homogeneous slab, $\alpha$ and $\beta$ are constant, and \eqref{C1-D1} and
\eqref{D101-def} -- \eqref{C2-1-def} give
	\begin{align}
	&C_1=\alpha, &&D_{1,0}=\beta-\frac{1}{\alpha}, && D_{1,1}=\frac{1}{\alpha},\nn\\
	&C_{2,0}=\frac{\alpha}{2}\left(\beta-\frac{1}{\alpha}\right), &&
	C_{2,1}=\frac{1}{2}.\nn
	\end{align}
Substituting these in \eqref{R1m1-def} -- \eqref{t22=}, we find
	\begin{align}
	&R_{1,-1}=T_{1,-1}=\frac{i}{2}\left(\beta-\frac{1}{\alpha}\right),
	&& R_{1,1}=-\frac{i(\alpha^2-1)}{2\alpha},\nn\\
	&T_{1,1}=\frac{i(\alpha-1)^2}{2\alpha},
	&&R^l_{2,0}=-\frac{1}{2\alpha}\left(\beta-\frac{1}{\alpha}\right),\nn\\
	&R^{l/r}_{2,-2}=T_{2,-2}=-\frac{1}{4}\left(\beta-\frac{1}{\alpha}\right)^2,
	&&R^l_{2,2}=\frac{\alpha^4-1}{4\alpha^2},\nn\\
	&R^r_{2,0}=\frac{2\alpha-1}{2\alpha}\left(\beta-\frac{1}{\alpha}\right),
	&&R^r_{2,2}=\frac{(\alpha^2-1)(\alpha^2-4\alpha+1)}{4\alpha^2},\nn\\
	&T_{2,0}=\frac{\alpha-1}{2\alpha}\left(\beta-\frac{1}{\alpha}\right),
	&&T_{2,2}=-\frac{(\alpha-1)^2(\alpha^2+1)}{4\alpha^2}.\nn
	\end{align}
These equation together with \eqref{rs=} and \eqref{ts=} imply
	\begin{align}
	R_1&=\frac{i}{2}\left[\left(\beta-\frac{1}{\alpha}\right)|\sec\theta|
	-\frac{(\alpha^2-1)|\!\cos\theta|}{\alpha}\right],
	\label{R1-slab=}\\
	R_2^l&=\frac{1}{4}\left[
	-\frac{2}{\alpha}\left(\beta-\frac{1}{\alpha}\right)
	-\left(\beta-\frac{1}{\alpha}\right)^2\sec^2\theta
	+\frac{(\alpha^4-1)\cos^2\theta}{\alpha^2}\right],\\
	R_2^r&=\frac{1}{4}\left[
	2\left(2-\frac{1}{\alpha}\right)\left(\beta-\frac{1}{\alpha}\right)
	-\left(\beta-\frac{1}{\alpha}\right)^2\sec^2\theta
	+\frac{(\alpha^2-1)(\alpha^2-4\alpha+1)\cos^2\theta}{\alpha^2}\right],\\
	T_1&=\frac{i}{2}\left[\left(\beta-\frac{1}{\alpha}\right)|\sec\theta|
	+\frac{(\alpha-1)^2|\!\cos\theta|}{\alpha}\right],\\
	T_2&=\frac{1}{4}\left[	\frac{2(\alpha-1)}{\alpha}\left(\beta-\frac{1}{\alpha}\right)
	-\left(\beta-\frac{1}{\alpha}\right)^2\sec^2\theta
	-\frac{(\alpha-1)^2(\alpha^2+1)\cos^2\theta}{\alpha^2}\right].
	\label{T2-slab=}
	\end{align}
Inserting these relations in \eqref{R=2} and \eqref{T=2}, we find the leading-order and next-to-leading-order terms in the low-frequency expansions of the reflection and transmission amplitudes. We have checked by explicit calculations that the resulting expressions coincide with those obtained by expanding the right-hand sides of (\ref{RR-homogen}) and (\ref{T-homogen}) with $a=0$ in powers of $k\ell$ and keeping the linear and quadratic terms.

\ed

\bibitem{dassios2000low}
G. Dassios, R. Kleinman, Low frequency scattering, Oxford University Press, 2000.

\bibitem{newton2013scattering} 
R. G. Newton, Scattering theory of waves and particles, Springer Science \& Business Media, 2013.

\bibitem{silfvast1996laser}
W. T. Silfvast, Laser Fundamentals, Cambridge University Press, 1996.

\bibitem{loran2021dynamical}
F. Loran, A. Mostafazadeh, Dynamical formulation of low-energy scattering in one dimension, J. of Math. Phys., 62, 4, 2021.

\bibitem{Faddeev1967}
L. D. Faddeev, The theory of the perturbation of continuous spectra for singular differential operators, AMS Translations, Series 2, 65, 139-166, 1967. 

\bibitem{yafaev1992mathematical}
D. R. Yafaev, Mathematical scattering theory: general theory, AMS, 105, 1992.

\bibitem{jones1941new}
R. C. Jones, A new calculus for the treatment of optical systems {i.} description and discussion of the calculus, JOSA, 31, 7, 488-493, 1941. 

\bibitem{abeles1950recherches}
F. Abel{\`e}s, Recherches sur la propagation des ondes {\'e}lectromagn{\'e}tiques sinuso{\"\i}dales dans les milieux stratifi{\'e}s-{A}pplication aux couches minces, Annal. de Phys., 12, 5, 596-640, 1950.

\bibitem{thomson1950transmission}
W. T. Thomson, Transmission of elastic waves through a stratified solid medium, Jour. of App. Phys., 21, 2, 89-93, 1950.

\bibitem{uzdin2012scattering}
R. Uzdin, N. Moiseyev, Scattering from a waveguide by cycling a non-{H}ermitian degeneracy, Phys. Rev. A: Atomic, Molecular, and Opt. Phys., 85, 3, 2012. 

\bibitem{ahmed2001schrodinger}
Z. Ahmed, Schr{\"o}dinger transmission through one-dimensional complex potentials, Phys. Rev. A, 64, 4, 2001. 

\bibitem{rayleigh1881x}
L. Rayleigh, X. On the electromagnetic theory of light, Lond. Edin. Dub. Phil. Mag. J. Sci., 12, 73, 81-101, 1881. 

\bibitem{ammari2000low} 
H. Ammari, J.-C. N{\'e}d{\'e}lec, Low-frequency electromagnetic scattering, SIAM J. Math. Anal., 31, 4, 836-861, 2000. 

\bibitem{bolle1987scattering}
D. Boll{\'e}, F. Gesztesy, M. Klaus, Scattering theory for one-dimensional systems with $\oint dx V (x)= 0$, J. Math. Anal. App., 122, 2, 496-518, 1987.

\bibitem{mostafazadeh2013invisibility}
A. Mostafazadeh, 
Invisibility and {PT} symmetry, 
Phys.~Rev.~A {\bf 87}, 012103  (2013). 

\bibitem{feng2013experimental}
L. Feng, Y.-L. Xu, W. S. Fegadolli, M.-H. Lu, J. E. B. Oliveira, V. R. Almeida, Y.-F. Chen, A. Scherer, Experimental demonstration of a unidirectional reflectionless parity-time metamaterial at optical frequencies, Nature Materials, 12, 2, 108-113, 2013. 

\bibitem{Southwell1983}
W. H. Southwell, Gradient-index antireflection coatings, Opt. Lett., 8, 11, 584-586, 1983. 

\bibitem{Poitras2004}
D. Poitras, J. A. Dobrowolski, Toward perfect antireflection coatings. 2. {T}heory, App. Opt., 43, 6, 1286-1295, 2004. 

\end{thebibliography}

\ed